# Trajectory Planning for Connected and Automated Vehicles at Isolated Signalized Intersections under Mixed Traffic Environment

Chengyuan Ma[a], Chunhui Yu[a, *], Xiaoguang Yang[a]

[a]*Key Laboratory of Road and Traffic Engineering of the Ministry of Education, Tongji University, 4800 Cao'an Road, Shanghai, P.R. China.*

**Abstract**

Trajectory planning for connected and automated vehicles (CAVs) has the potential to improve operational efficiency and vehicle fuel economy in traffic systems. Despite abundant studies in this research area, most of them only consider trajectory planning in the longitudinal dimension or assume the fully CAV environment. This study proposes an approach to the decentralized planning of CAV trajectories at an isolated signalized intersection under the mixed traffic environment, which consists of connected and human-driven vehicles (CHVs) and CAVs. A bi-level optimization model is formulated based on discrete time to optimize the trajectory of a single CAV in both the longitudinal and lateral dimensions given signal timings and the trajectory information of surrounding vehicles. The upper-level model optimizes lateral lane-changing strategies. The lower-level model optimizes longitudinal acceleration profiles based on the lane-changing strategies from the upper-level model. Minimization of vehicle delay, fuel consumption, and lane-changing costs are considered in the objective functions. A Lane-Changing Strategy Tree (LCST) and a Parallel Monte-Carlo Tree Search (PMCTS) algorithm are designed to solve the bi-level optimization model. CAV trajectories are planned one by one according to their distance to the stop bar. A rolling horizon scheme is applied for the dynamic implementation of the proposed model with time-varying traffic condition. Numerical studies validate the advantages of the proposed trajectory planning model compared with the benchmark cases without CAV trajectory planning. The average fuel consumption and lane-changing numbers of CAVs can be reduced noticeably, especially with high traffic demand. The average delay of CAVs is reduced by ~2 s on average, which is limited due to the fixed signal timing plans. The trajectory planning of CAVs also reduces the delay and the fuel consumption of CHVs and the mixed traffic, especially with high penetration rates of CAVs. The sensitivity analysis shows that the control zone length of 200 m is sufficient to ensure the satisfactory performance of proposed model.

*Keywords:* mixed traffic environment; trajectory planning; longitudinal and lateral trajectory; bi-level optimization model

* Corresponding author. Tel.: (+86) 166-2118-2873
  *Email address:* hughyu90@tongji.edu.cn

# 1 Introduction

Recent advances of connected and autonomous vehicle (CAV) technologies are regarded as one of the most promising solutions to improve traffic safety and efficiency, which have become a major topic of concern for policymakers and researchers. The connected vehicle technology enables real-time communication between vehicles (V2V) and between vehicles and infrastructures (V2I); and the automated vehicle technology enables precise control of vehicle trajectories. The combination of connected vehicle and automated vehicle technologies further enables the trajectory planning for CAVs and offers new approaches to traffic operations.

By designing CAV trajectories, a number of existing studies aim to improve traffic operational performance such as safety, efficiency, and environmental friendliness at particular facilities, such as to reduce environmental impacts along highway segments (Lu et al., 2019) and to coordinate vehicles going through the highway merging area (Hu and Sun, 2019). At urban intersections, vehicles arriving during red lights may stop at stop bars and then accelerate when traffic lights turn green. This process raises the travel time and fuel consumption of vehicles as well as reduces intersection capacity. By appropriate trajectory planning, CAVs can slow down in advance to avoid stops and queues at a stop bar (Feng et al., 2018; He et al., 2015). Therefore, CAV trajectory planning at intersections is widely investigated for both optimizing vehicle driving behaviors and improving traffic flow performances.

For the energy saving and emission reduction of vehicles at intersections, CAV trajectory planning is adopted in the development of eco-drive systems. Eco-driving systems usually provide ecological speed profiles to a vehicle based on the predicted behaviors of its preceding vehicles in a look forward horizon and traffic signal timings (Xia et al., 2013; He et al., 2015, Yang et al., 2017). The current states of the preceding vehicles can be received through vehicle-to-x (V2X) communication (Hu et al., 2016), or detected by the on-board sensors of CAVs (Kamal et al., 2015) and loop detectors (Jiang et al., 2017). And their future trajectories are predicted using longitudinal driving behavior models, such as Gipps car-following model (Kamal et al., 2015) and the intelligent driver model (Jiang et al., 2017). The optimal speed profiles of the target vehicle are usually generated in an optimal control framework to reduce fuel consumption and improve mobility and comfort, under the constraints of traffic rules (He et al., 2015; Hu et al., 2016; Jiang et al., 2017).

In addition, CAV trajectory planning is introduced in the research area of CAV-based traffic control at signalized and "signal-free" intersections to mitigate congestion, lessen the risk of crashes, and reduce fuel consumption and emissions under both fully and partially CAV environment. In the fully CAV environment, the conflicts between vehicles with incompatible movements at intersections can be avoided by controlling CAV trajectories without explicit traffic lights (Dresner and Stone, 2008; Mirheli et al., 2019). A planned trajectory strategy could lead a CAV to slow down in advance to avoid stops and queues at stop bars for the elimination of start-up lost time and the improvement of driving experience at signalized intersections (He et al., 2015; Feng et al., 2018; Yu et al., 2018; Zhang and Cassandras, 2019; Yu et al., 2019; Kamal et al., 2020). However, the fully CAV environment cannot be realized in the near future. It is widely expected that the mixed traffic with human-driven vehicles (HVs) and autonomous vehicles (AVs) will exist in the next 20–30 years (Zheng et al., 2020). Compared with the fully CAV environment, the trajectory planning in the partially CAV environment needs to consider the driving behaviors of HVs, whose trajectories cannot be precisely controlled directly. Most of related studies focus on the optimization of the longitudinal speed profiles of CAVs based on predicted future trajectories of HVs. (Yang et al., 2016; Elefteriadou et al., 2017; Zhao et al., 2018; Ghiasi et al., 2019; Yao and Li, 2020). Yang et al. (2016) used kinematic wave theory to predict the queue length of HVs at a signalized intersection and then proposed a bi-level model to optimize both signal timings and CAV longitudinal trajectories. Pourmehrab et al. (2017) proposed an Intelligent Intersection Control System (IICS) for mixed traffic flows at signalized intersections. HV trajectories were first predicted by the Gipps Car-Following model and CAV trajectories were then optimized for minimum delay. Guo et al. (2019) proposed an efficient dynamic programming with shooting heuristic (DP-SH) algorithm for the integrated optimization of CAV trajectories and signal timings. HV trajectories were predicted based on the entry information from vehicle detectors upstream of the intersection at the beginning of the trajectory control section. Zhao et al. (2018) proposed a model predictive control (MPC) method to minimize the fuel consumption for platoons of mixed CAVs and HVs passing a signalized intersection. Yao and Li (2020) proposed a decentralized control model for CAV trajectory planning at a signalized intersection with a single-lane road to minimize the travel time, fuel consumption, and safety risks of each CAV. The results showed that the decentralized model overperformed the benchmark centralized control model in terms of computational efficiency without significant loss of the system optimality. Above studies have also validated that CAV trajectory planning can not only improve the driving experience of CAVs but also influence their following vehicles and optimize the overall traffic operational performances.

However, the trajectory planning methods in these studies usually assume no lane changing and only optimize longitudinal trajectories with the consideration of other vehicles in the same lane. The lack of considering lateral trajectories (i.e., lane-changing behaviors) makes these studies inapplicable in the real world because mandatory lane changing is inevitable at urban intersections. Although studies in the research area on automatic vehicle control have investigated two-dimensional trajectory planning problems, they usually focus on the design of precise geometry properties of trajectories of individual vehicles in a short horizon (e.g., 10 s) based on local traffic environment (González et al., 2016). The trajectory planning for multiple vehicles considering global traffic information in a long horizon is missing. It is worth mentioning that the developed MILP model in Yu et al. (2018) optimized both longitudinal and lateral vehicle trajectories at isolated intersections. All vehicles were assumed to enter the control zone in dedicated lanes. That is, only optional lane changing was considered. In addition, the approach is confined to the fully CAV environment. Vehicle trajectory planning in both longitudinal and lateral dimensions under the partially CAV environment remains to be investigated.

Several challenges emerge in the two-dimensional trajectory planning for CAVs at intersections. Firstly, rather than just consider the car-following relationship with the vehicles in the same lane, more factors like vehicles in the other lanes should be also well concerned. Secondly, the lane-changing strategy should cooperate with the longitudinal speed profile, because a CAV's lane-changing maneuvers will affect the solution space of the longitudinal trajectory and the car-following strategy affects the lateral trajectory in return. Thirdly, computational burden may render real-time implementation difficult, especially with high traffic demand. This challenge is highlighted in several pioneering studies on centralized optimization frameworks due to the complex nature of multi-trajectory planning problems (Li and Li, 2019). One approach is using discrete time in model formulation (Miyatake et al., 2011; Li and Li, 2019) for the application of numerical solution algorithms (e.g., dynamic programming). In addition, a vehicle trajectory is divided into several segments as an approximation to reduce the dimension of solution space (Zhou et al.2017; Ma et al., 2017; Feng et al., 2018). And heuristic algorithms are also investigated (Guo et al., 2019). Different from centralized optimization, decentralized schemes allow vehicles to negotiate with each other and plan their driving strategies themselves (Malikopoulos et al., 2018; Liu et al., 2018; Mirheli et al., 2019; Yao and Li, 2020), which shows advantages on computation efficiency and is more suitable for real-time applications.

Realizing the research gaps, this study presents an approach to the optimization of CAV trajectories in a decentralized way at isolated signalized intersections under the mixed traffic environment, which consists of connected and human-driven vehicles (CHVs) and CAVs. All vehicles are assumed to be connected in this study as governments tend to vigorously promote the V2X technologies for their benefits on safety, mobility, and environmental friendliness. For example, National Highway Traffic Safety Administration (NHTSA) is going to require vehicle-to-vehicle (V2V) capability for all new vehicles (U.S. Department of Transportation, 2017), and the connectivity adoption rate is expected to reach 100% in 2023 according to the corresponding analysis report (U.S. Department of Transportation, 2016). A bi-level optimization model is formulated based on discrete time to optimize longitudinal and lateral trajectories of a single CAV given signal timings and predicted/planned trajectories of other CHVs and CAVs. The upper-level model optimizes the lateral trajectory, (i.e., lane choices). The lower-level model optimizes the longitudinal trajectory (i.e., acceleration profiles) based on the lane-changing strategies from the upper-level model. The objective is to minimize vehicle delay, fuel consumption, and lane-changing costs. A Parallel Monte-Carlo Tree Search (PMCTS) algorithm is applied to solve the bi-level optimization model. CAV trajectories are planned one by one according to their distance to the stop bar in a decentralized way. A rolling horizon implementation procedure is designed for the application of the proposed model to time-varying traffic condition.

The remainder of this paper is organized as follows. Section 2 describes the addressed problem. Section 3 formulates the bi-level model of CAV trajectory optimization under the mixed traffic condition. Section 4 designs the solution algorithms for the bi-level model and the implementation procedure with time-varying traffic condition. Section 5 conducts numerical studies and sensitivity analysis. Finally, Section 6 delivers the conclusions and future research directions.

## 2 Problem description and notations

### 2.1 Problem description

**Fig. 1** shows the details of one approach arm of a typical signalized intersection with four arms. Each approach lane is dedicated to a specific vehicle movement. There is a no-changing zone close to the stop bar, where lane changing is not allowed. Vehicles have to finish lane changing before the no-changing zone, which is the current practice in the real world. CAVs and CHVs coexist in the approach lanes and both follow the signals at the intersection. Vehicles are connected within the control zone. And their real-time states $s^\omega(t)$ (i.e., lane choice $k^\omega(t)$, location $x^\omega(t)$, speed $v^\omega(t)$, and acceleration rate $a^\omega(t)$) are assumed to be collected and shared without communication delay. The trajectory of a vehicle can then be captured by $s^\omega = \{s^\omega(t) | t = t_0^\omega, t_0^\omega + 1, \ldots, t_0^\omega + h\}$. $t_0^\omega$ and $h$ are the start time step and the length of the trajectory planning horizon, respectively. Based on the online collected vehicle states and signal timings, the trajectories of CAVs within the control zone are dynamically planned to reduce their delay and fuel consumption. For the convenience of modeling, vehicle trajectory $s^\omega$ is decomposed into a lateral lane-changing strategy and a longitudinal acceleration profile. A **Lane-Changing Gap (LCG)** is defined as the spatial interval between two adjacent vehicles in the same lane at a time step, e.g., the marked gap $g_{\omega_f}^{\omega_p}(t)$ between preceding vehicle $\omega_p$ and following vehicle $\omega_f$ at time step $t$ in **Fig. 1**. Then the lane-changing strategy $\boldsymbol{g}^\omega$ of vehicle $\omega$ can be represented by the LCG choice $g^\omega(t)$ at each time step, i.e., $\boldsymbol{g}^\omega = \{g^\omega(t) | t = t_0^\omega, t_0^\omega + 1, \ldots, t_0^\omega + h\}$. Note that the current lane choice $k^\omega(t_0^\omega)$ of vehicle $\omega$ is denoted as the LCG between its current preceding and following vehicles. For example, vehicle $\omega_0$ can change to lane 1 by taking LCG $g_{\omega_f}^{\omega_p}(t)$ at time step $t$. Such LCGs always exist in each approach lane by placing virtual vehicles at the beginning and ending locations of the lane, whose longitudinal locations are set as $-M$ and $M$, respectively. Similarly, the longitudinal acceleration profile is denoted as $\boldsymbol{a}^\omega = \{a^\omega(t) | t = t_0^\omega, t_0^\omega + 1, \ldots, t_0^\omega + h\}$. In this way, $s^\omega$ can also be represented as $s^\omega = (\boldsymbol{g}^\omega, \boldsymbol{a}^\omega) = \{(g^\omega(t), a^\omega(t)) | t = t_0^\omega, t_0^\omega + 1, \ldots, t_0^\omega + h\}$.

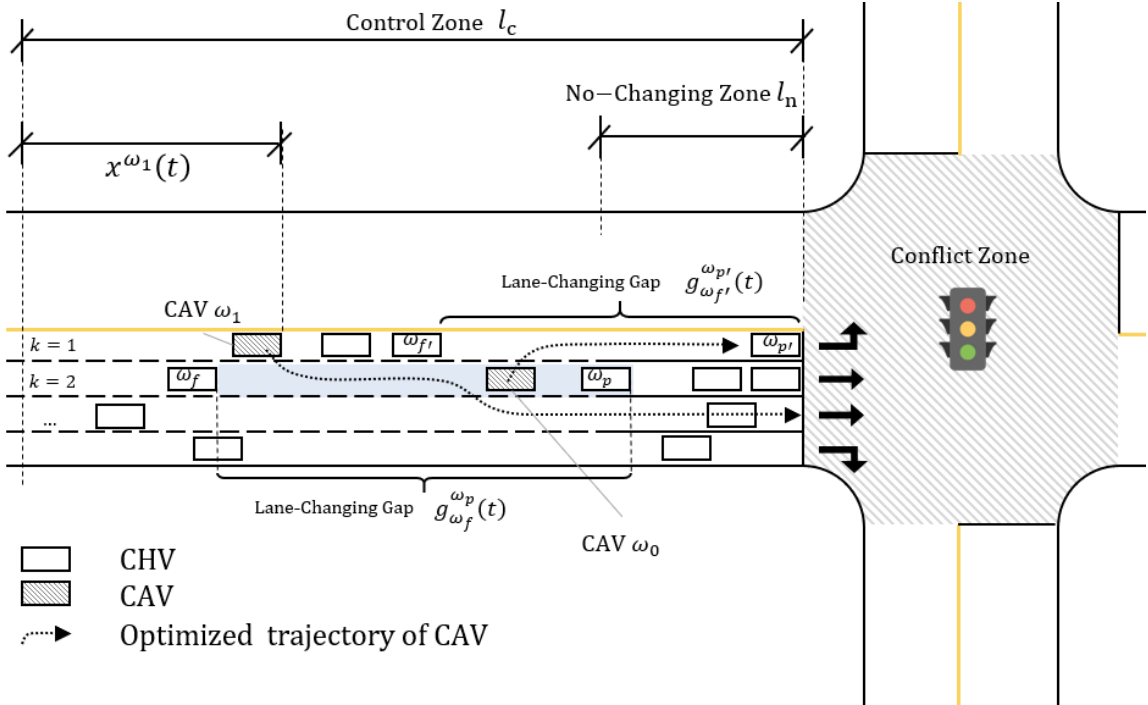

**Fig. 1** Trajectory planning for CAVs under the mixed traffic environment.

For computational efficiency, CAV trajectories are optimized in a decentralized way. CAV trajectories are optimized one by one according to their longitudinal location in the approach lanes. Each CAV $\omega$ collects the information on the signal timings and the current states of the vehicles within the control zone as well as the planned trajectories of its preceding CAVs ($\mathbf{\Omega}_A^\omega$). The preceding vehicles $\mathbf{\Omega}^\omega = \mathbf{\Omega}_H^\omega \cup \mathbf{\Omega}_A^\omega$, where $\mathbf{\Omega}_H^\omega$ denotes the preceding CHVs, are the ones closer to the stop bar than CAV $\omega$ regardless of their occupied lanes, i.e., $\mathbf{\Omega}^\omega = \{\omega^* | x_{\omega^*}(t_0^\omega) \geq x_\omega(t_0^\omega)\}$. Based on the collected real-time data, CAV $\omega$ dynamically optimizes its trajectory for mobility and fuel efficiency with time-varying traffic condition.

In the trajectory planning of CAV $\omega$, the surrounding vehicles are divided into three categories, namely, the preceding CAVs $\mathbf{\Omega}_A^\omega$, the preceding CHVs $\mathbf{\Omega}_H^\omega$, and the following vehicles $\bar{\mathbf{\Omega}}^\omega$. CAV $\omega$ knows the future trajectories of the preceding CAVs because their planned trajectories are shared. The future trajectories of the preceding CHVs and the following vehicles are predicted with the aid of car-following and lane-changing models. For the concerns of fairness, CAV $\omega$ optimizes its trajectory in the way that the trajectories of its preceding vehicles are not affected based on the planned and predicted trajectories of its preceding CAVs and CHVs, respectively. That is, the planned trajectory of CAV $\omega$ does not force its preceding CAVs or CHVs to change their trajectories, which has the potential to reduce the negative impacts on the traffic flow. In contrast, the following vehicles may be forced to decelerate or change lanes for safety concerns, for example, when CAV $\omega$ changes lanes. As CHVs may not follow their predicted trajectories, a rolling horizon scheme is applied to cater to time-varying traffic condition, similar to the philosophy in MPC (Camacho,2004).

*2.2 Notations*

Main notations applied hereafter are summarized in the following **Table 1**.

**Table 1** Notations.

| **General notations** | |
|---|---|
| M | A sufficiently large number |
| $t_0^\omega$ | Start time step of the trajectory planning horizon of CAV $\omega$ |
| $k$ | Approach lane index |
| $\omega$ | Vehicle index |
| $\mathbf{\Omega}$ | Set of the vehicles in the approaching lanes in the considered arm at $t_0^\omega$ |
| $\mathbf{\Omega}^\omega$ | Set of the preceding vehicles of vehicle $\omega$ at $t_0^\omega$; the subsets of CHVs and CAVs are denoted as $\mathbf{\Omega}_H^\omega$ and $\mathbf{\Omega}_A^\omega$, respectively (i.e., $\mathbf{\Omega}^\omega = \mathbf{\Omega}_H^\omega \cup \mathbf{\Omega}_A^\omega$) |
| $\bar{\mathbf{\Omega}}^\omega$ | Set of the following vehicles of vehicle $\omega$ at $t_0^\omega$, i.e., $\bar{\mathbf{\Omega}}^\omega = \mathbf{\Omega} \setminus (\mathbf{\Omega}^\omega \cup \{\omega\})$ |
| $g_{\omega_f}^{\omega_p}(t)$ | LCG between preceding vehicle $\omega_p$ and following vehicle $\omega_f$ at time step $t$ |
| $V_p(\cdot)$ | Preceding vehicle of an LCG |
| $V_f(\cdot)$ | Following vehicle of an LCG |
| $x_p(\cdot)$ | Longitudinal location of the preceding vehicle of an LCG, m |
| $x_f(\cdot)$ | Longitudinal location of the following vehicle of an LCG, m |
| $k(\cdot)$ | Lane occupied by the preceding and the following vehicles of an LCG |
| $\mathbf{K}^\omega$ | Set of the dedicated approach lanes for the movement of vehicle $\omega$ |
| $\mathbf{G}^\omega(t)$ | Set of the feasible LCGs of vehicle $\omega$ at time step $t$ |
| $\mathbf{L}^\omega$ | Set of the feasible lane-changing strategies of vehicle $\omega$ |
| **Parameters** | |
| $h$ | Length of the trajectory planning horizon in time steps |
| $\Delta t$ | Length of a time step, s |
| $l_c$ | Length of the control zone, m |
| $l_n$ | Length of the no-changing zone, m |
| $l^\omega$ | Length of vehicle $\omega$, m |
| $\alpha^\omega$ | $\alpha^\omega = 1$, if $\omega$ is a virtual vehicle; $\alpha^\omega = 0$, otherwise |
| $d_p$ | Safety distance to the new preceding vehicle after lane changing, m |
| $d_f$ | Safety distance to the new following vehicle after lane changing, m |
| $\tau_h$ | Redundant time steps of the trajectory planning horizon for computational feasibility |
| $\tau_{lc}$ | Minimum time interval between two lane-changing behaviors of a single vehicle, s |
| $\tau_{cf}$ | Time displacement in Newell's car-following model, s |
| $d_{cf}$ | Space displacement in Newell's car-following model, m |
| $a_L^\omega$ | Maximum deceleration rate of vehicle $\omega$, m²/s |
| $a_U^\omega$ | Maximum acceleration rate of vehicle $\omega$, m²/s |
| $v_U$ | Speed limit in the approach lanes, m/s |
| $v_U^c$ | Speed limit within the conflict zone, m/s |
| $r^k(t)$ | $r^k(t) = 1$, if the traffic light for approach lane $k$ is red at time step $t$; 0, otherwise |
| **Decision variables in lane-changing strategy optimization** | |
| $g^\omega(t)$ | LCG taken by CAV $\omega$ at time step $t$ |
| $\boldsymbol{g}^\omega$ | Lane-changing strategy of CAV $\omega$; $\boldsymbol{g}^\omega = \{g^\omega(t) | t = t_0^\omega, t_0^\omega + 1, \dots, t_0^\omega + h\}$ |
| **Decision variables in acceleration profile optimization** | |
| $a^\omega(t)$ | Acceleration rate of CAV $\omega$ at time step $t$, m²/s |
| $\boldsymbol{a}^\omega$ | Acceleration profile of CAV $\omega$; $\boldsymbol{a}^\omega = \{a^\omega(t) | t = t_0^\omega, t_0^\omega + 1, \dots, t_0^\omega + h\}$ |
| **Auxiliary variables** | |
| $\vartheta^\omega(t)$ | Binary variable. $\vartheta^\omega(t) = 1$, if vehicle $\omega$ changes lanes at time step $t$; 0, otherwise |
| $k^\omega(t)$ | Index of the lane taken by vehicle $\omega$ at time step $t$ |
| $\delta^\omega(t)$ | Binary variable. $\delta^\omega(t) = 0$, if vehicle $\omega$ has passed the stop bar by time step $t$; 1, otherwise |
| $t_c^\omega$ | Time step within which vehicle $\omega$ passes the stop bar |
| $x^\omega(t)$ | Distance between vehicle $\omega$ and the start of the control zone at time step $t$, m |
| $v^\omega(t)$ | Speed of vehicle $\omega$ at time $t$, m/s |
| $s^\omega(t)$ | State of vehicle $\omega$ at time step $t$; $s^\omega(t) = \{x^\omega(t), v^\omega(t), a^\omega(t), k^\omega(t)\}$ |
| $\boldsymbol{s}^\omega$ | Trajectory of vehicle $\omega$; $\boldsymbol{s}^\omega = \{s^\omega(t) | t = t_0^\omega, t_0^\omega + 1, \dots, t_0^\omega + h\}$ |

## 3 CAV trajectory optimization model

*3.1 Model framework*

**Fig. 2** shows the bi-level optimization framework of the trajectory planning for CAV $\omega$. At the start time step $t_0^\omega$ of the planning horizon, CAV $\omega$ collects the information on the signal timings, the initial states $s^\omega(t_0^\omega)$ and movement directions of the vehicles within the control zone ($\mathbf{\Omega}$), and the planned trajectories of its preceding CAVs ($\mathbf{\Omega}_A^\omega$).

At the initialization stage, the trajectories of the preceding CHVs ($\mathbf{\Omega}_H^\omega$) is firstly predicted with the aid of the second-order car-following model in Eissfeldt (2004) and a modified lane-changing model based on the one in Erdmann (2014) when the preceding CAVs ($\mathbf{\Omega}_A^\omega$) keep their planned trajectories. Then the initial feasible trajectories of CAV $\omega$ and its following vehicles ($\bar{\mathbf{\Omega}}^\omega$) are generated using the same car-following and lane-changing models on the condition that the trajectories of the preceding vehicles are not affected. The generated trajectory of CAV $\omega$ serves as the initial feasible solution of its trajectory planning. And the planned/predicted trajectories of the other vehicles serve in the safety constraints of the proposed model. The modified lane-changing model is described in Section 4.1.

In the bi-level model of the trajectory planning, the lane-changing strategy $\boldsymbol{g}^\omega$ and the longitudinal acceleration profile $\boldsymbol{a}^\omega$ are jointly optimized. In the upper-level model, $\boldsymbol{g}^\omega$ and $\boldsymbol{a}^\omega$ are optimized to minimize the overall cost including travel time, fuel consumption, and lane-changing cost. Due to the complex relationship between $\boldsymbol{g}^\omega$ and $\boldsymbol{a}^\omega$, the lower-level optimization model is formulated to determine the optimal $\boldsymbol{a}^\omega$ for minimum travel time and fuel consumption given $\boldsymbol{g}^\omega$ from the upper-level model. In turn, $\boldsymbol{g}^\omega$ and $\boldsymbol{a}^\omega$ determine the objective function of the upper-level model. The outputs of the bi-level optimization model are the optimal lane-changing strategy $\boldsymbol{g}^\omega$ and acceleration profile $\boldsymbol{a}^\omega$ (i.e., trajectory $\boldsymbol{s}^\omega$). Note that in the trajectory planning, $\boldsymbol{g}^\omega$ and $\boldsymbol{a}^\omega$ are constrained to have no impacts on the planned/predicted trajectories of the preceding vehicles. They are also constrained to guarantee safety between CAV $\omega$ and its following vehicles considering the trajectory adjustment of the following vehicles.

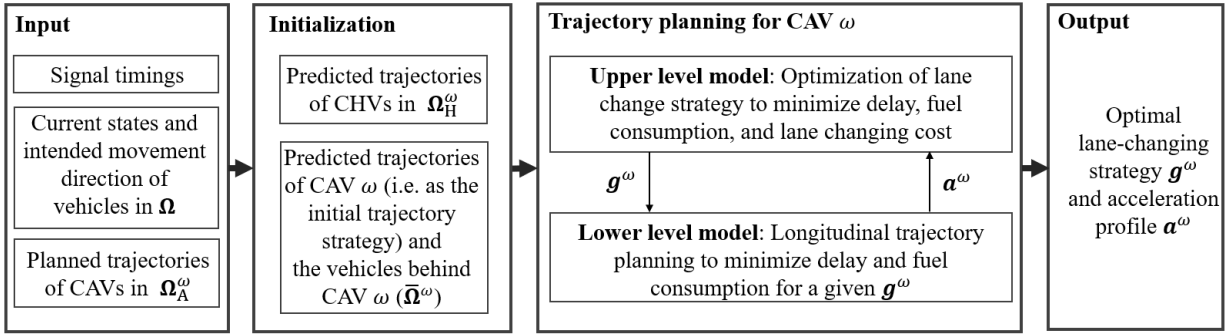

**Fig. 2** Model framework of trajectory planning.

*3.2 Upper-level model: lane-changing strategy optimization*

The upper-level optimization model is formulated in **Eq**s. (1)–(6):
**P1:**
$$\min_{\boldsymbol{g}^\omega} C[\boldsymbol{a}^\omega, \boldsymbol{g}^\omega] \tag{1}$$

s.t.
$$g^\omega(t_0^\omega) = g_{\omega_0^f}^{\omega_0^p}(t_0^\omega) \tag{2}$$

$$g^\omega(t) \in \mathbf{G}^\omega(t) = \left\{ g_{\omega_2}^{\omega_1}(t) \middle| \mathbf{F}_1\left(g_{\omega_2}^{\omega_1}(t)\right) \le \mathbf{0}, \omega_1 \in \mathbf{\Omega}^\omega, \omega_2 \in \mathbf{\Omega} \right\}, \forall t = t_0^\omega, t_0^\omega + 1, \dots, t_0^\omega + h \tag{3}$$

$$\mathbf{F}_2(\boldsymbol{g}^\omega) \le \mathbf{0} \tag{4}$$

$$k\left(g^\omega(t_0^\omega + h)\right) \in \mathbf{K}^\omega \tag{5}$$

$$\boldsymbol{a}^\omega = \boldsymbol{h}\left(\boldsymbol{g}^\omega; s^\omega(t_0^\omega)\right) \tag{6}$$

**Eq.** (1) is the objective function of minimizing the overall cost considering travel time, fuel consumption, and lane-changing cost. It is formulated as

$$C[\boldsymbol{a}^\omega, \boldsymbol{g}^\omega] = \alpha_1(t_c^\omega - t_0^\omega)\Delta t + \alpha_2 \sum_{t=t_0^\omega}^{t_c^\omega} |a^\omega(t)| + \alpha_3 \sum_{t=t_0^\omega}^{t_0^\omega+h} \vartheta^\omega(t) \qquad (7)$$

where $t_c^\omega$ is the time step within which CAV $\omega$ passes the stop bar and $(t_c^\omega - t_0^\omega)\Delta t$ is regarded as the travel time. Since the free-flow travel time is fixed, the minimization of travel time is equivalent to the minimization of vehicle delay, which is adopted as the primary objective in the proposed model. The secondary objective is the minimization of $|a^\omega(t)|$ to improve the smoothness of the longitudinal trajectory, which also helps reduce fuel consumption (Feng, et al., 2018). The tertiary objective is the minimization of the cost of the lane-changing strategy $\boldsymbol{g}^\omega$, i.e., the lane-changing number $\sum_{t=t_0^\omega}^{t_0^\omega+h} \vartheta^\omega(t)$. $\vartheta^\omega(t) = 1$, if CAV $\omega$ changes lanes at time step $t$; $\vartheta^\omega(t) = 0$, otherwise. The minimization of $\sum_{t=t_0^\omega}^{t_0^\omega+h} \vartheta^\omega(t)$ improves the smoothness of the lateral trajectory. To differentiate the primary, the secondary, and the tertiary objectives, their weighting parameters, i.e., $\alpha_1$, $\alpha_2$, and $\alpha_3$, should satisfy $\alpha_1 \gg \alpha_2 \gg \alpha_3 > 0$.

Note that $t_c^\omega$ is a decision variable and the second term $\sum_{t=t_0^\omega}^{t_c^\omega} |a^\omega(t)|$ in **Eq. (7)** makes **P1** difficult to be solved. To handle this problem, $\sum_{t=t_0^\omega}^{t_c^\omega} |a^\omega(t)|$ is reformulated as

$$\sum_{t=t_0^\omega}^{t_c^\omega} |a^\omega(t)| = \sum_{t=t_0^\omega}^{t_0^\omega+h} \delta^\omega(t)|a^\omega(t)| \qquad (8)$$

where $\delta^\omega(t)$ is an auxiliary variable. $\delta^\omega(t) = 0$, if CAV $\omega$ has passed the stop bar by time step $t$; $\delta^\omega(t) = 1$, otherwise. Note that **Eq. (8)** can be easily linearized. Similarly, the first term $(t_c^\omega - t_0^\omega)\Delta t$ in **Eq. (7)** is reformulated as

$$(t_c^\omega - t_0^\omega)\Delta t = \sum_{t=t_0^\omega}^{t_0^\omega+h} \delta^\omega(t)\Delta t \qquad (9)$$

**Eq. (2)** indicates that the initial taken LCG is $g_{\omega_0^f}^{\omega_0^p}(t_0^\omega)$ at the start time step $t_0^\omega$ of the planning horizon, where $\omega_0^p$ and $\omega_0^f$ are the exact preceding and following vehicles of CAV $\omega$ in the same lane, respectively. **Eq. (3)** guarantees that the taken LCG should be in the feasible LCG set $\mathbf{G}^\omega(t)$ at each time step $t$ considering the planned/predicted trajectories of the other vehicles. The details of the constraints $\mathbf{F}_1(\cdot) \leq \mathbf{0}$, which define $\mathbf{G}^\omega(t)$, are described in Section 3.2.1. **Eq. (4)** defines the constraints of the sequential lane-changing maneuvers of CAV $\omega$, which are described in Section 3.2.2. **Eq. (3)** and **Eq. (4)** determine the solution space $\mathbf{L}^\omega$ of lane-changing strategies $\boldsymbol{g}^\omega$. **Eq. (5)** guarantees that CAV $\omega$ passes the stop bar in the dedicated lanes for its movement (i.e., in left-turning, through, or right-turning lanes). **Eq. (6)** indicates that the acceleration profile $\boldsymbol{a}^\omega$ is determined by the lane-changing strategy $\boldsymbol{g}^\omega$ and the initial state $s^\omega(t_0^\omega)$, which is formulated as the lower-level optimization model. In this way, $\boldsymbol{g}^\omega = \{g^\omega(t) | t = t_0^\omega, t_0^\omega + 1, \ldots, t_0^\omega + h\}$ is regarded as the only decision variables in the upper-level optimization model.

### 3.2.1 Set of feasible LCGs

At each time step $t$, the available LCGs of CAV $\omega$ are known given the planned/predicted trajectories of the other vehicles in the control zone. For safety concerns, only partial LCGs are feasible, the set of which is denoted as $\mathbf{G}^\omega(t)$. CAV $\omega$ may collide with the preceding or the following vehicle of the LCG if it takes an infeasible LCG. The constraints $\mathbf{F}_1\left(g_{\omega_2}^{\omega_1}(t)\right) \leq \mathbf{0}$ defining $\mathbf{G}^\omega(t)$ in **Eq. (3)** are expressed as

$$d_p + d_f \leq x_p\left(g_{\omega_2}^{\omega_1}(t)\right) - x_f\left(g_{\omega_2}^{\omega_1}(t)\right) \qquad (10)$$

where $x_p\left(g_{\omega_2}^{\omega_1}(t)\right)$ and $x_f\left(g_{\omega_2}^{\omega_1}(t)\right)$ are the longitudinal locations of the preceding and the following vehicles of LCG $g_{\omega_2}^{\omega_1}(t)$, respectively. **Eq. (10)** indicates that the spacing of a feasible LCG in $\mathbf{G}^\omega(t)$ should be sufficiently large to avoid collisions. $d_p$ (or $d_f$) is the safety distance between CAV $\omega$ and the preceding (or the following) vehicle of LCG $g_{\omega_2}^{\omega_1}(t)$. In fact, the values of $d_p$ and $d_f$ are related to the states (e.g., speeds and acceleration rates) of the preceding and the following vehicles of LCG $g_{\omega_2}^{\omega_1}(t)$ as well as CAV $\omega$ (Zheng, 2014). The exact modeling of such longitudinal safety constraints is presented in Section 3.3.2. In **Eq.** (10), $d_p$ and $d_f$ take fixed values in a conservative way. That is, an LCG in $\mathbf{G}^{\omega}(t)$ may not satisfy the longitudinal safety constraints in the lower-level model in Section 3.3.2. The purpose of introducing **Eq.** (10) is to reduce the solution space of $g^{\omega}(t)$ in the upper-level model.

### 3.2.2   Sequential lane-changing maneuvers

**Eq.** (4) describes the constraints of the sequential lane-changing maneuvers of CAV $\omega$. The constraints $\mathbf{F}_2(\boldsymbol{g}^{\omega}) \leq \mathbf{0}$ include **Eq**s. (11)–(12):

$$-\mathrm{M}\big(2 - \vartheta^{\omega}(t_1) - \vartheta^{\omega}(t_2)\big) + \tau_{\mathrm{lc}} \leq (t_2 - t_1)\Delta t$$
$$\forall t_2 = t_1 + 1, \ldots, t_0^{\omega} + h; t_1 = t_0^{\omega}, t_0^{\omega} + 1, \ldots, t_0^{\omega} + h \tag{11}$$

$$\big|k\big(g^{\omega}(t+1)\big) - k\big(g^{\omega}(t)\big)\big| = \vartheta^{\omega}(t+1), \forall t = t_0^{\omega}, t_0^{\omega} + 1, \ldots, t_0^{\omega} + h - 1 \tag{12}$$

where $\tau_{\mathrm{lc}}$ is the minimum time interval between two consecutive lane-changing maneuvers; $k\big(g^{\omega}(t)\big)$ is the lane occupied by the preceding and the following vehicles of LCG $g^{\omega}(t)$. **Eq.** (11) guarantees the minimum time interval between two consecutive lane-changing maneuvers for safety concerns. **Eq.** (12) indicates that CAV $\omega$ can change at most one lane in one lane-changing maneuver (i.e., $\big|k\big(g^{\omega}(t+1)\big) - k\big(g^{\omega}(t)\big)\big| \leq 1$). It also guarantees that $\vartheta^{\omega}(t) = 1$, if CAV $\omega$ changes lanes (i.e., $k\big(g^{\omega}(t+1)\big) \neq k\big(g^{\omega}(t)\big)$); $\vartheta^{\omega}(t) = 0$, otherwise (i.e., $k\big(g^{\omega}(t+1)\big) = k\big(g^{\omega}(t)\big)$). Instant lane-changing maneuvers are assumed for simplicity. Note that $g^{\omega}(t+1)$ may differ from $g^{\omega}(t)$ when CAV $\omega$ remains in the same lane. Because the preceding or the following vehicle of CAV $\omega$ may vary with time-varying traffic condition.

### 3.3   Lower-level model: acceleration profile optimization

The lower-level model optimizes the longitudinal acceleration profile $\boldsymbol{a}^{\omega} = \{a^{\omega}(t)|t = t_0^{\omega}, t_0^{\omega} + 1, \ldots, t_0^{\omega} + h\}$ of CAV $\omega$ given a lane-changing strategy $\boldsymbol{g}^{\omega}$ from the upper-level model. The objective is the minimization of travel time and fuel consumption, i.e., the first and the second components in **Eq.** (7):
**P2:**

$$\min_{\boldsymbol{a}^{\omega}} \alpha_1 \sum_{t=t_0^{\omega}}^{t_0^{\omega}+h} \delta^{\omega}(t)\Delta t + \alpha_2 \sum_{t=t_0^{\omega}}^{t_0^{\omega}+h} \delta^{\omega}(t)|a^{\omega}(t)| \tag{13}$$

The constraints are described in detail in Subsection 3.3.1-3.3.3.

### 3.3.1   Longitudinal vehicle kinematics

At the start of the planning horizon, the initial state $s^{\omega}(t_0^{\omega}) = \{x^{\omega}(t_0^{\omega}), v^{\omega}(t_0^{\omega}), a^{\omega}(t_0^{\omega}), k^{\omega}(t_0^{\omega})\}$ of CAV $\omega$ is known. The second-order vehicle kinematics model is applied:

$$v^{\omega}(t+1) = v^{\omega}(t) + \Delta t \times a^{\omega}(t), \forall t = t_0^{\omega}, t_0^{\omega} + 1, \ldots, t_0^{\omega} + h - 1 \tag{14}$$

$$x^{\omega}(t+1) = x^{\omega}(t) + \frac{\Delta t}{2} \times \big(v^{\omega}(t) + v^{\omega}(t+1)\big), \forall t = t_0^{\omega}, t_0^{\omega} + 1, \ldots, t_0^{\omega} + h - 1 \tag{15}$$

$$-a_L^{\omega} \leq a^{\omega}(t) \leq a_U^{\omega}, \forall t = t_0^{\omega}, t_0^{\omega} + 1, \ldots, t_0^{\omega} + h \tag{16}$$

$$0 \leq v^{\omega}(t) \leq v_U, \forall t = t_0^{\omega}, t_0^{\omega} + 1, \ldots, t_0^{\omega} + h \tag{17}$$

where $a_L^{\omega}$ and $a_U^{\omega}$ are the absolute values of the maximum deceleration and acceleration rates, respectively. $v_U$ is the speed limit in the approach lanes. Generally, the speed limit $v_U^c$ within the conflict zone is lower than that in the approach lanes. Additional constraints are applied:

$$v_U^c + \mathrm{M}\delta(t) \geq v^{\omega}(t), \forall t = t_0^{\omega}, t_0^{\omega} + 1, \ldots, t_0^{\omega} + h \tag{18}$$

**Eq**. (18) indicates that $v^{\omega}(t) \leq v_U^c$ after CAV $\omega$ passes the stop bar (i.e., $\delta(t) = 0$).

### 3.3.2 Longitudinal safety

When CAV $\omega$ travels in an approach lane, it should keep a safe distance from its preceding/following vehicle in the same lane. The longitudinal safety constraints include:

$$x^\omega(t) \leq x^{\omega'}\left(t - \frac{\tau_{cf}}{\Delta t}\right) - d_{cf}, \forall \omega' = V_p(g^\omega(t)); t = t_0^\omega + \frac{\tau_{cf}}{\Delta t}, t_0^\omega + \frac{\tau_{cf}}{\Delta t} + 1, \ldots, t_0^\omega + h \qquad (19)$$

$$x^\omega(t) \leq l_c + M\left(3 - \alpha^{\omega'} - r^k(t) - \delta^\omega(t-1)\right),$$
$$\forall \omega' = V_p(g^\omega(t)); k = k(g^\omega(t)); t = t_0^\omega + 1, t_0^\omega + 2, \ldots, t_0^\omega + h \qquad (20)$$

$$x^\omega(t) \geq x^{\omega'}\left(t + \frac{\tau_{cf}}{\Delta t}\right) + d_{cf}, \forall \omega' = V_f(g^\omega(t)) \in \Omega^\omega; t = t_0^\omega, t_0^\omega + 1, \ldots, t_0^\omega + h - \frac{\tau_{cf}}{\Delta t} \qquad (21)$$

$$x^\omega(t) \geq x^{\omega'}(t) + \frac{\left(v^{\omega'}(t)\right)^2}{2a_L^{\omega'}} - M(1 - \vartheta^\omega(t)), \forall \omega' = V_f(g^\omega(t)) \in \overline{\Omega}^\omega; t = t_0^\omega, t_0^\omega + 1, \ldots, t_0^\omega + h \qquad (22)$$

where $V_p(g^\omega(t))$ and $V_f(g^\omega(t))$ are the preceding and the following vehicles of LCG $g^\omega(t)$, respectively. Newell's car-following model (Newell, 2002) is applied in **Eq**. (19) to guarantee the safety distance between CAV $\omega$ and its preceding vehicle $\omega'$ in the same lane. $\tau_{cf}$ and $d_{cf}$ are the time and space displacement in Newell's car-following model, respectively. $\Delta t$ should be selected properly to make $\tau_{cf}/\Delta t$ an integer. To model the impacts of red lights, **Eq**. (20) is introduced. If CAV $\omega$ has passed the stop bar by time step $t - 1$, i.e., $\delta^\omega(t-1) = 0$, the traffic light has no influence on its trajectory at time step $t$. Otherwise, when the traffic light is red ($r^k(t) = 1$) and the preceding vehicle is the virtual one ($\alpha^{\omega'} = 1$), **Eq**. (19) and **Eq**. (20) guarantee $x^\omega(t) \leq l_c$ for CAV $\omega$. **Eq**. (21) guarantees the safety distance between CAV $\omega$ and its following vehicle $\omega' \in \Omega^\omega$ in the same lane. In **Eq**. (19) and **Eq**. (21), the trajectories of the preceding and the following vehicles $\omega'$ are known and fixed. And **Eq**. (21) indicates that the trajectory planning for CAV $\omega$ does not affect the planned/predicted trajectories of vehicle $\omega' \in \Omega^\omega$.

**Eq**. (22) guarantees the safety distance between CAV $\omega$ and its immediate following vehicle $\omega' \in \overline{\Omega}^\omega$ in the target lane if CAV $\omega$ changes lanes, i.e., $\vartheta^\omega(t) = 1$. Different from **Eq**. (21), following vehicle $\omega'$ may be forced by CAV $\omega$ to decelerate to guarantee the safety distance after CAV $\omega$ changes lanes. The location $x^{\omega'}(t)$ and the speed $v^{\omega'}(t)$ at each time step are estimated based on the predicted trajectory of vehicle $\omega'$. Note that the estimation of $x^{\omega'}(t)$ and $v^{\omega'}(t)$ in **Eq**. (22) may not be accurate after the trajectory of vehicle $\omega'$ is influenced by that of CAV $\omega$. Once vehicle $\omega'$ is blocked by CAV $\omega$ and decelerates for safety, its trajectory may differ from the predicted one. To handle the inaccurate modeling of these safety constraints, a rolling horizon scheme is proposed in Section 4.3.

Additionally, the following constraints **Eq**. (23) is applied to ensure that CAV $\omega$ finishes lane-changing maneuvers outside the no-changing zone for safety concerns:

$$x^\omega(t) - M(1 - \vartheta^\omega(t)) \leq l_c - l_n, \forall t = t_0^\omega, t_0^\omega + 1, \ldots, t_0^\omega + h \qquad (23)$$

where $l_c$ is the length of the control zone; $l_n$ is the length of the no-changing zone. After CAV $\omega$ enters the no-changing zone (i.e., $x^\omega(t) > l_c - l_n$), **Eq**. (23) guarantees $\vartheta^\omega(t) = 0$.

### 3.3.3 Indicator of passing the stop bar

$\delta^\omega(t)$ is defined to indicate whether CAV $\omega$ has passed the stop bar by time step $t$. $\delta^\omega(t) = 0$, if CAV $\omega$ has passed the stop bar by time step $t$ (i.e., $x^\omega(t) > l_c$); and $\delta^\omega(t) = 1$, otherwise. This is specified by

$$M(1 - \delta^\omega(t)) \geq x^\omega(t) - (l_c + \epsilon) \geq -M\delta^\omega(t), \forall t = t_0^\omega, t_0^\omega + 1, \ldots, t_0^\omega + h \qquad (24)$$

where $\epsilon$ is a small number to handle the strictly less-than constraints $x^\omega(t) > l_c$. By the end of the planning horizon, CAV $\omega$ is expected to have passed the stop bar for a completely planned trajectory:

$$x^\omega(t_0^\omega + h) \geq l_c + l^\omega + \epsilon \qquad (25)$$

## 4 Solution algorithms

*4.1 Trajectory prediction algorithm*

The future trajectories of the other vehicles determine the solution space of CAV $\omega$'s trajectory planning. Therefore, in the initialization process in **Fig. 2**, we need to predict the future trajectories of the vehicles in $\mathbf{\Omega}_H^\omega$ during the planning horizon, and generate the initial feasible trajectories for CAV $\omega$ and the vehicles in $\bar{\mathbf{\Omega}}^\omega$. The trajectory prediction is based on current traffic conditions and vehicle driving behaviors determined by car-following and lane-changing models. The process of the trajectory prediction follows:

**Step 1**: Initialize the prediction start time $t = t_0^\omega$ and the current state of each vehicle

**Step 2**: Obtain the states (i.e., lane choices, acceleration rates, speeds, and locations) of CAVs in $\mathbf{\Omega}_A^\omega$ at time step $t + 1$ according to their planned trajectories.

**Step 3**: Predict the states of CHVs in $\mathbf{\Omega}_H^\omega$ at time step $t + 1$ one by one according to their longitudinal locations in the approach lanes. For a certain CHV in $\mathbf{\Omega}_H^\omega$, the lane-changing decision at time step $t + 1$ is first predicted based on the lane-changing model in Erdmann (2014) with several additional rules to guarantee CAVs in $\mathbf{\Omega}_A^\omega$ can keep their planned trajectories. These additional rules include: 1) CHVs in $\mathbf{\Omega}_H^\omega$ are not allowed to overtake their preceding CAVs for the concerns of fairness. That is, CAVs in $\mathbf{\Omega}_A^\omega$ will not be blocked in their lane by their following CHVs in $\mathbf{\Omega}_H^\omega$. 2) CHVs in $\mathbf{\Omega}_H^\omega$ will decelerate or change lanes for helping the CAVs in $\mathbf{\Omega}_A^\omega$ lane changing towards their lane, as they seeing the turn-signals of the preceding vehicles.

If the CHV is predicted to change lanes at time step $t + 1$, its longitudinal state is then determined simultaneously by the lane-changing model to promote the successful execution of the desired lane-changing maneuver. Otherwise, the corresponding longitudinal state at time step $t + 1$ is determined by the car-following model in Eissfeldt (2004).

**Step 4**: Predict the states of CAV $\omega$ and the vehicles in $\bar{\mathbf{\Omega}}^\omega$ at time step $t + 1$. The process is similar to the trajectory prediction in **Step 3**. And the same two rules are also adopted in the trajectory prediction of vehicles in $\bar{\mathbf{\Omega}}^\omega$ and CAV $\omega$ to guarantee that the planned or predicted trajectories of the vehicles in $\mathbf{\Omega}^\omega$ are not be affected. Because of the uncertainty nature of the traffic dynamics, a rolling horizon scheme is proposed in Section 4.3 to handle varying traffic conditions.

**Step 5**: If CAV $\omega$ has not passed the stop bar at time step $t + 1$, then set $t = t + 1$ and go to **Step 2**. Otherwise, set the trajectory planning horizon length as $h = (t + 1) - t_0^\omega + \tau_h$, and output the predicted trajectories.

The predicted travel time of CAV $\omega$ in the approach lanes (i.e., $(t + 1) - t_0^\omega$) is used to determine the length of its trajectory planning horizon. According to the constraints in Section 3.3.3, the trajectory planning horizon length should be sufficiently long to guarantee CAV $\omega$ to pass the stop bar. Otherwise, the proposed optimization model becomes infeasible. However, the model complexity and thus the computational burden increase with the increasing planning horizon length. Note that the minimization of travel time is the primary objective in **Eq**. (7) and the predicted initial trajectory of CAV $\omega$ provides an upper bound of the minimal travel time. For the trade-off between model feasibility and computational efficiency, the length of the trajectory planning is set as $h = (t + 1) - t_0^\omega + \tau_h$. $\tau_h$ is the redundant time steps considering the inaccurate modeling of vehicle driving behaviors and the uncertainty in traffic environment.

*4.2 Solution algorithm for the bi-level optimization model*

In the upper-level model, LCG choices $g^\omega(t)$ are the decision variables, which are discrete. In the lower-level model, acceleration rates $a^\omega(t)$ are the decision variables together with auxiliary binary variables. Given a lane-changing strategy $\boldsymbol{g}^\omega$ from the upper-level model, the lower-level model is an MILP model. The difficulty in solving the bi-level optimization model lies in the large solution space of lane-changing strategies (i.e., $\mathbf{L}^\omega$), especially with a long planning horizon and a large number of vehicles in the control zone. In this section, a Lane-Changing Strategy Tree (LCST) and a PMCTS algorithm are designed for solutions. The process of solving the bi-level optimization model follows:

**Step 1**: At time step $t_0^\omega$, CAV $\omega$ collects the states of the vehicles within the control zone, the planned trajectories of the CAVs in $\mathbf{\Omega}_A^\omega$, and the signal timing plans.

**Step 2**: Predict the trajectories of the vehicles in $\mathbf{\Omega}_H^\omega$ and $\bar{\mathbf{\Omega}}^\omega$ and get the initial trajectory of CAV $\omega$ according to the algorithm in Section 4.1. The initial trajectory is set as the current best solution $\boldsymbol{s}_o^\omega$ in the trajectory planning and the horizon length $h$ is determined at the same time.

**Step 3**: Generate the feasible LCG set $\mathbf{G}^\omega(t)$ at each time step in the planning horizon based on the planned trajectories of the vehicles in $\mathbf{\Omega}_A^\omega$ and the predicted trajectories of the vehicles in $\mathbf{\Omega}_H^\omega$ and $\bar{\mathbf{\Omega}}^\omega$.

**Step 4**: Generate the feasible lane-changing strategy set $\mathbf{L}^\omega$ by constructing the LCST with the tree breadth-first search algorithm in Section 4.2.1.

**Step 5**: Search for the optimal lane-changing strategy with the PMCTS algorithm in Section 4.2.2. The performance of each lane-changing strategy is evaluated by the objective function **Eq. (1)** after the lower-level model (**P2**) is solved.

**Step 6**: Output the optimal trajectory of CAV $\omega$ including the lane-changing strategy and the longitudinal acceleration profile.

*4.2.1   Lane-Changing Strategy Tree*

**Fig. 3** illustrates the construction of the LCST to generate the feasible lane-changing strategy set $\mathbf{L}^\omega$. The nodes in each layer denote the feasible LCGs at each time step that satisfy constraints **Eq.** (3). The red node in the first layer is the root node of the LCST denoting the taken LCG in lane $k$ at the start time step $t_0^\omega$ of the planning horizon (i.e., constraint **Eq.** (2)). The edge connecting two nodes in adjacent layers denotes the sequential lane-changing maneuvers at two adjacent time steps, which satisfies constraints **Eq.** (4). The solid lines indicate remaining in the same lane and the dash-dotted lines indicate changing to an adjacent lane. A path from the root node to a leaf node (e.g., the yellow node in **Fig. 3**) in the lane satisfying constraints **Eq.** (5) denotes a feasible lane-changing strategy $\boldsymbol{g}^\omega$. The tree breadth-first search algorithm (**Algorithm 1**) is applied to search for all feasible paths satisfying constraints **Eq**s. (2)–(5). And the lane-changing strategy set $\mathbf{L}^\omega$ is presented by the constructed LCST.

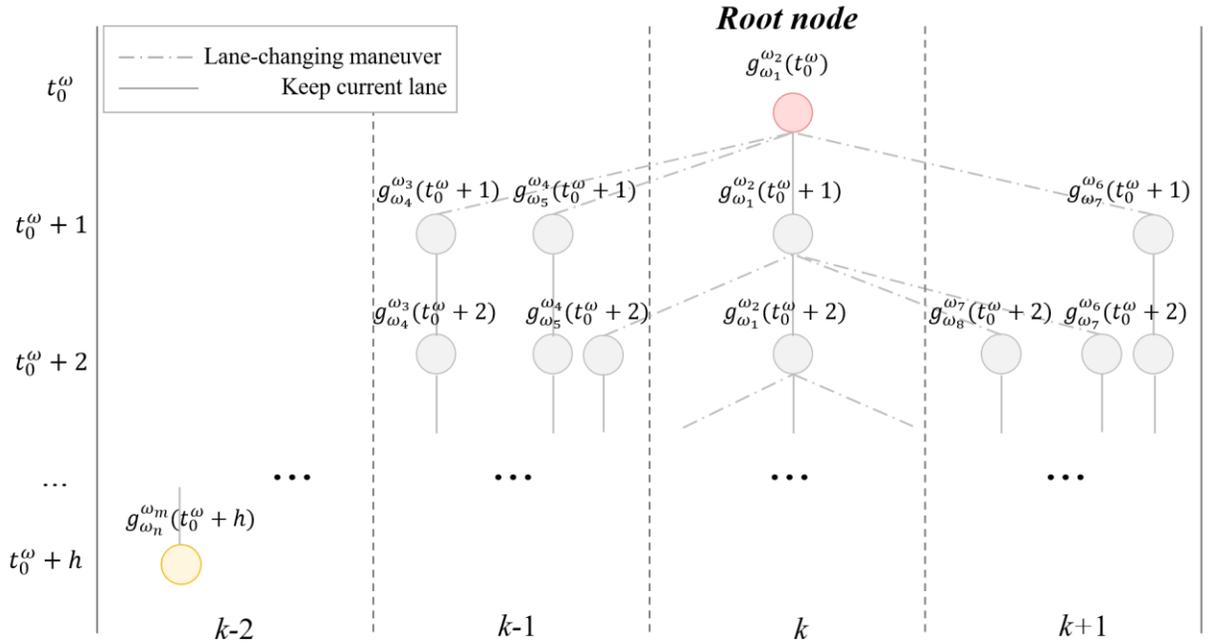

**Fig. 3** Illustration of the LCST.

**Algorithm 1** Tree breadth-first search algorithm.

**Input**: $\mathbf{G}^\omega(t)$, $t_0^\omega$, $h$, $g^\omega(t_0^\omega)$

**Output**: LCST (i.e., $\mathbf{L}^\omega$)

Initialize the root node of the LCST using $g^\omega(t_0^\omega)$;
Set $t = t_0^\omega$;
**while** $t < t_0^\omega + h$ **do**
  **while** $\mathbf{G}^\omega(t+1)$ is not empty **then**
    Set $g_{t+1} = \mathbf{G}^\omega(t+1).\text{pop}()$.
    **Foreach** node $g_t$ in layer $t$
      If $g_{t+1}$ and $g_t$ satisfy constraints **Eqs.** (11)–(12) **then**
        Add $g_{t+1}$ to layer $t+1$ as a child node of $g_t$. If $g$ is added multiple times, it means
        the added node $g$ is connected to multiple father nodes (e.g., $g_{\omega_5}^{\omega_4}(t_0^\omega + 2)$ in **Fig. 3**).
        In this way, one father node can have multiple child nodes but one child node only has
        one father node.
  Set $t = t + 1$.
Remove the paths whose leaf node does not satisfy constraints **Eq.** (5).
The generated LCST represents the lane-changing strategy set $\mathbf{L}^\omega$ and each path represents a feasible lane-changing strategy.

### 4.2.2 Parallel Monte-Carlo Tree Search Algorithm

Monte-Carlo Tree Search has caught attention due to its outstanding performance in the field of Computer Go (Kocsis et al., 2006). In this study, it is applied to the search of the constructed LCST for sub-optimal solutions within limited time. Each node in the LCST has a reward value and the algorithm tends to explore the nodes with higher reward values. And the reward values are updated as the algorithm goes. The reward value $R(g)$ of node $g$ is formulated as (Kocsis et al., 2006):

$$R(g) = e^{-c_1 f(g)} + c_2 \sqrt{\frac{1}{N(g)}} \tag{26}$$

In **Eq.** (26), the first term is the "performance component", which represents the best performance of all the explored paths (i.e., lane-changing strategies) that contain node $n$. $f(g)$ is updated each time node $g$ is explored. The nodes with higher performance are more likely to lead to the solutions with high quality. The second term is the "exploration component". $N(g)$ denotes the exploration numbers of node $g$. In this way, the Monte-Carlo Tree Search algorithm prefers to explore the nodes with fewer exploration numbers to avoid falling into local optimal. $c_1$ and $c_2$ are positive constant parameters to balance exploration and exploitation. The detailed algorithm follows:

**Step 1 Initialization:** Set the initial trajectory of CAV $\omega$, which is generated by the algorithm in Section 4.1, as the current best trajectory solution $\mathbf{s}_0^\omega$. Initialize the reward value $R(g)$ of each node $g$ as: 1) $f(g) = C_0$ for all nodes in the LCST, where $C_0$ is the cost (i.e., the objective function **Eq.** (7)) of the initial trajectory solution $\mathbf{s}_0^\omega$, 2) $N(g) = 1$ for all nodes in the LCST. That is, the initial reward value of each node is $R(g) = e^{-c_1 C_0} + c_2$.

**Step 2 Selection**: Select a lane-changing strategy $\mathbf{g}^\omega$ by traversing from the root node to a leaf node. When there are multiple child nodes at a father node $g$, the one with the highest reward value is selected. The path representing the selected lane-changing strategy $\mathbf{g}^\omega$ is then pruned from the LCST to avoid repeated selection in the **Selection** step.

**Step 3 Simulation:** The longitudinal trajectory optimization model (**P2**) with the selected $\mathbf{g}^\omega$ is built as an MILP model and is solved by a solver. If the corresponding **P2** is feasible, get the optimal trajectory strategy $\mathbf{s}^\omega$ with the cost $C^*$ and go **Step 4**. Otherwise, go to **Step 5**.

**Step 4 Backpropagation:** Updates the reward values $R(g)$ of the nodes in the selected path via

$$f(g) = \min(f(g), C^*) \tag{27}$$

$$N(g) = N(g) + 1 \tag{28}$$

If solution $s^\omega$ is better than the current best trajectory solution $s_0^\omega$, then set $s_0^\omega = s^\omega$.

**Step 5** If either the time limit is reached or all the paths in the LCST are pruned, output the current best trajectory solution $s_0^\omega$. Otherwise, go to **Step 2**.

Note that constructing and solving the MILP models (**P2**) for different lane-changing strategies are independent from each other. Therefore, parallel computing techniques can be applied for computational efficiency. A PMCTS algorithm is designed as shown in **Fig. 4**. The **Selection**, **Simulation**, and **Backpropagation** steps can be conducted in multiple threads. A crucial problem in the parallelization is using mutexes to prevent data corruption (Chaslot et al., 2008). In this study, a mutex is used to locks the global LCST in the **Selection** and **Backpropagation** steps. Only one thread can access the LCST at one time.

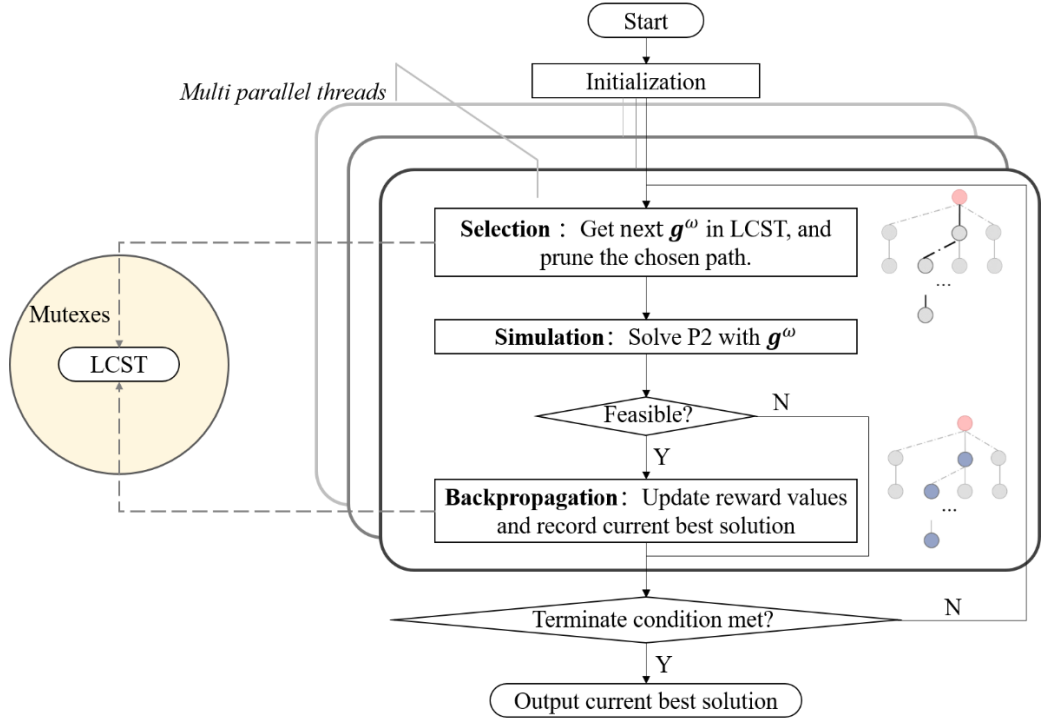

**Fig. 4** Scheme of the PMCTS algorithm.

*4.3 Rolling horizon scheme*

A rolling horizon scheme is proposed for the dynamic implementation of the trajectory optimization model to cater to varying traffic conditions, which is shown in **Fig. 5**. The trajectory planning for CAVs in the approach lanes is executed in a decentralized way at each time step. CAV trajectories are planned one by one according to their distance to the stop bar. The trajectory planning procedure for one CAV $\omega$ follows the algorithm in Section 4.2.

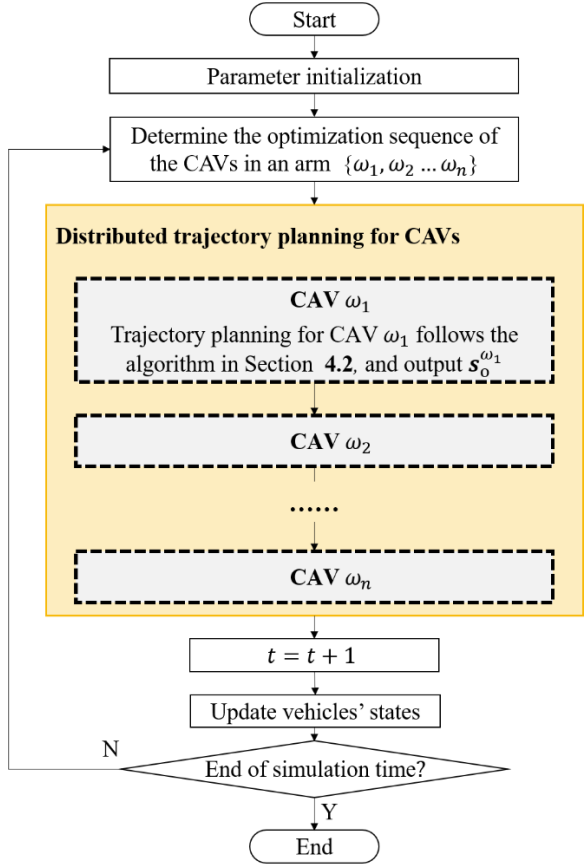

**Fig. 5** Rolling horizon scheme.

## 5 Numerical studies

### 5.1 Experimental data

A micro-simulation of the typical intersection in **Fig. 1** is applied to explore the benefits of the proposed trajectory optimization model. Since CAV trajectory planning in one arm is independent of that in another arm, one arm of the intersection is taken for the experiments, which has four approach lanes including a left-turning lane, two through lanes, and a right-turning lane. The length of the control zone is $l_c = 500$ m. The length of the no-changing zone is $l_n = 30$ m. The speed limits in the control zone and in the conflict zone are $v_U = 16.6$ m/s and $v_U^c = 10$ m/s, respectively.

Both CAVs and CHVs take the same vehicle size and the same parameters in driving behavior models to eliminate the impacts of the difference in driving behaviors for the better illustration of the benefits of the proposed trajectory planning. The vehicle length $l^\omega = 4$ m. The time displacement and the space displacement in the car-following model are $\tau_{cf} = 1$ s and $d_{cf} = 6$ m. The minimum time interval between two lane-changing behaviors is $\tau_{lc} = 5$ s. The absolute values of the maximum acceleration and deceleration rates are $a_U^\omega = 2$ m/s² and $a_L^\omega = 4$ m/s². The safety distances in **Eq.** (10) in the lane-changing planning for CAVs are $d_p = 5$ m and $d_f = 6$ m. The weighting parameters $\alpha_1$, $\alpha_2$, and $\alpha_3$ in the objective function **Eq.** (7) are 1000 $s^{-1}$, 10 $s^2/m$, and 1. The time step length is $\Delta t = 1$ s and the redundant time steps of the planning horizon are $\tau_h = 5$ (i.e., 5 s).

Five levels of traffic demand are tested in simulation as shown in **Table 1**. Both under- and over-saturated traffic are included. Vehicle arrivals conform to Poisson distribution and vehicles enter the control zone in a random lane. The cycle length of the used signal plan is 60 s. Left-turning and through vehicles share the same phase with the green duration of 27 s. The yellow time is 3 s. CAVs are allowed to pass the stop bar within the first second of the yellow time in the trajectory planning and the last two seconds are seen as the red time for safety concerns. CHVs follow the current rules in practice. When they come across a yellow light, they would decelerate if they could stop safely at the stop bar. Otherwise, they would pass the stop bar during the yellow light. Right-turning vehicles are not controlled by signals.

**Table 1** Traffic demand.

| Demand level | Average arrival rate (pcu/h (v/c)) | | |
| --- | --- | --- | --- |
| | Through | Left-turning | Right-turning |
| 1 | 563 (0.25) | 253 (0.25) | 506 (0.25) |
| 2 | 1125 (0.50) | 506 (0.50) | 1012 (0.50) |
| 3 | 1688 (0.75) | 759 (0.75) | 1518 (0.75) |
| 4 | 2250 (1.00) | 1012 (1.00) | 2024 (1.00) |
| 5 | 2813 (1.25) | 1265 (1.25) | 2530 (1.25) |

The proposed algorithms are implemented in C#. The MILP model (**P2**) at the lower level is solved using Gurobi 9.0 (Gurobi Optimization Inc. 2019). The simulation is conducted in SUMO (Simulation of Urban MObility) (Krajzewicz, et al., 2012) on a desktop with an Intel 3.60 GHz 8 core CPU and 16 GB memory. Five random seeds are used in the simulation for each demand level considering stochastic vehicle arrivals. Each simulation run is 1800 s with a warm-up period of 150 s.

*5.2 Results and discussions*

*5.2.1 Computational efficiency*

As described in Section 4.2, the computational efficiency of the proposed trajectory planning algorithm, in which the PMCTS algorithm is embedded, can be improved by using parallel computing techniques with multiple threads. **Fig. 6** shows the average computational time of the trajectory planning algorithm at the second demand level (v/c = 0.5) with different numbers of available threads. When only one thread is available, the average computational time is more than 4.5 s. When the thread number increases to seven, the computational time decreases significantly to ~1 s. However, the improvement is insignificant when the number of available threads further increases. In that case, the computational efficiency is bounded by solving the MILP model (**P2**) in Step 3 of the PMCTS algorithm in each thread.

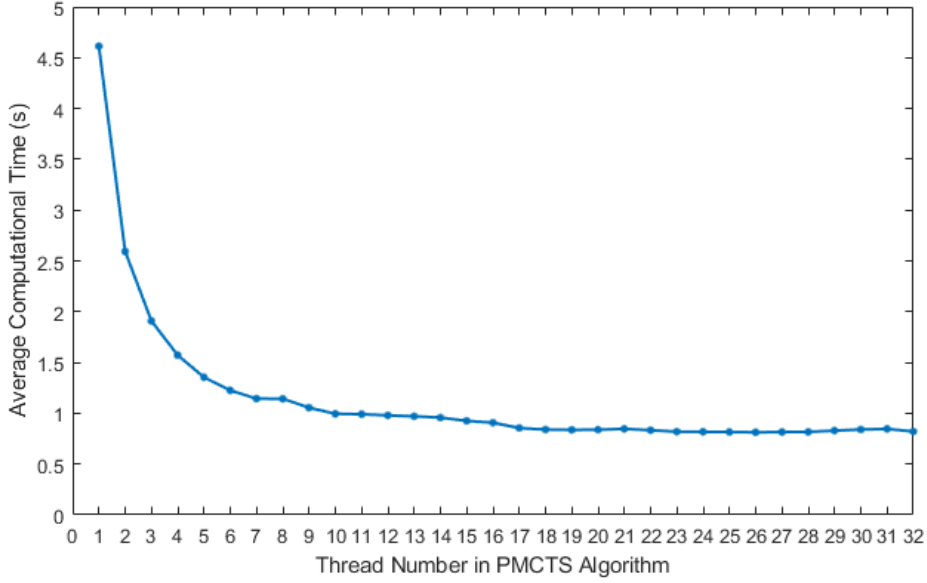

**Fig. 6** Computational efficiency of PMCTS algorithm.

For practice in field, a time limit can be set for the PMCTS algorithm to get a sub-optimal solution for the real-time implementation of the trajectory planning algorithm. **Table 2** shows the average optimality of sub-optimal solutions with different time limits and thread numbers. The solution optimality is measured by $\left(C(s_*^\omega) - C(s_o^\omega)\right)/\left(C(s^\omega) - C(s_0^\omega)\right) \times 100\%$, where $C(s_*^\omega)$, $C(s^\omega)$, and $C(s_0^\omega)$ are the costs (i.e., the objective function **Eq.** (7)) of the optimal solution, the sub-optimal solution, and the initial solution that is generated in Step 2 of the trajectory planning algorithm. When a sub-optimal solution $s^\omega$ has high solution quality, the optimality index is close to 100%. Otherwise, the optimality index is noticeably less than 100%. **Table 2** shows that a larger number of threads has the potential to find sub-optimal solutions of higher quality within a time limit. And a longer time limit helps lead to higher-quality solutions as well. When there are 16 threads, the average optimality index of the sub-optimal solutions can reach 95.0% within the time limit of 1.00 s, which shows the promising implementation of the trajectory planning algorithm in field.

**Table 2** Solution optimality with different time limits.

| Time limits (s) | Maximum Threads | | | | |
|---|---|---|---|---|---|
| | 1 | 4 | 8 | 16 | 32 |
| 0.50 | 16.0% | 28.3% | 31.0% | 40.0% | 56.0% |
| 0.75 | 26.1% | 34.6% | 52.0% | 72.0% | 81.0% |
| 1.00 | 26.2% | 39.0% | 75.0% | 95.0% | 97.0% |
| 1.25 | 27.4% | 59.0% | 91.0% | 96.0% | 98.0% |
| 1.50 | 30.2% | 75.3% | 92.5% | 96.9% | 98.4% |

One challenge of solving the proposed bi-level optimization model may lie in the increasing solution space with increasing traffic, which is known as the curse of dimensionality. **Table 3** shows the average computational time of the proposed trajectory planning algorithm at different traffic demand levels when there are 16 threads and no time limits. When the traffic demand increases fivefold from level 1 to level 5, the average number of vehicles in approach lanes increases by more than 12 times due to the congestion. However, the average number of lane-changing strategies in the solution space (i.e., the paths in the LCST) only increases by ~1.5 times. And the average computational time increases from 0.83 s to 2.17 s by ~1.5 times. The reason is that the number of feasible LCGs does not increase in proportion to the increase of traffic demand because of the small space headway with heavy traffic. Therefore, the trajectory planning algorithm is capable of handling both low and high traffic in terms of computational efficiency.

Table 3 Computational time at different traffic demand levels.

| Demand level | Average number of vehicles in approach lanes | Average number of lane-changing strategies in LCST | Average computational time (s) |
| --- | --- | --- | --- |
| 1 | 12 | 76 | 0.83 |
| 2 | 28 | 84.6 | 0.91 |
| 3 | 57 | 143.5 | 1.47 |
| 4 | 99 | 173.6 | 1.96 |
| 5 | 161 | 192.4 | 2.17 |

To fully explore the benefits of the proposed model, 16 threads and no time limits are used in the following studies.

*5.2.2  Benefits for CAVs*

The proposed trajectory planning model helps CAVs reduce delay and lane-changing numbers as well as raise fuel economy. **Fig. 7**, **Fig. 9**, and **Fig. 10** show the benefits of the planned trajectory strategies of CAVs compared with the benchmark trajectory strategies, which are generated without CAV trajectory planning in SUMO. Left-turning (L), right-turning (R), and through (T) CAVs are illustrated separately in **Fig. 7**(b), **Fig. 9**(b), and **Fig. 10**(b). The experiments are conducted under mixed traffic with the CAV penetration rate of 40%. The travel delay is calculated as the difference between the actual travel time and the free-flow travel time at the intersection, which is the sum of the travel times in the approaching lanes, within the intersection area, and in the exit lanes. The fuel consumption model in (Karsten et al., 2013) is applied for the estimation of fuel economy.

**Fig. 7**(a) shows the average delays of CAVs at various traffic demand levels. Compared with the benchmark trajectories, the optimized trajectories can reduce the travel delays of CAVs by ~5% at all the demand levels. **Fig. 7**(b) shows the average delays grouped by CAV movements. The delay reduction is more noticeable with high demand (v/c=1.0 and 1.25), which can reach ~2 s. Because CHVs stop at the stop bar more frequently with high demand and the optimized trajectories can help CAVs avoid such stops as shown in **Fig. 8**(a). As a result, CAVs can reduce the start-up loss time and the green time is fully utilized. Further, CAVs can have higher speeds passing the intersection due to the avoidance of stops at the stop bar. These also explain why the reduced delay of right-turning CAVs is less significant than those of left-turning and through CAVs as shown in **Fig. 7**(b). However, the reduced delay is limited because the signal timings are fixed.

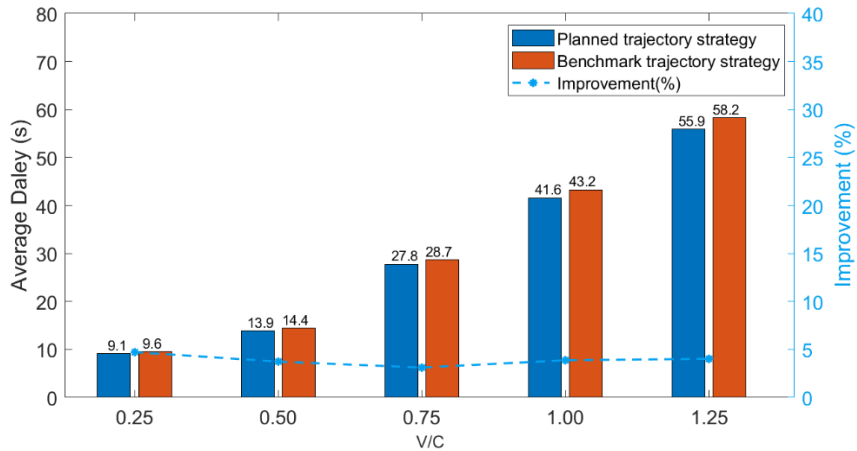

(a) CAV delay.

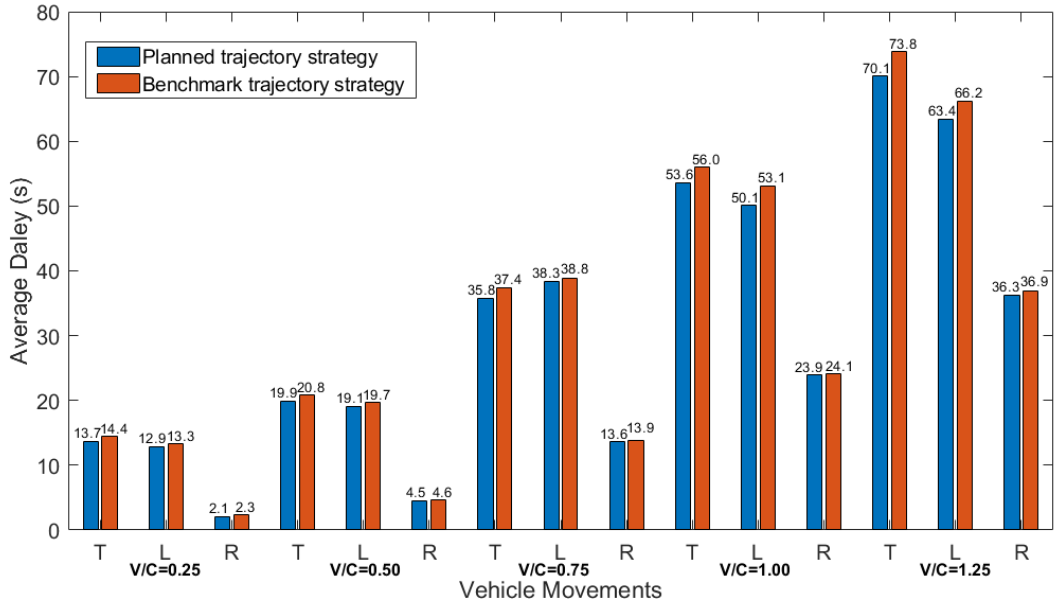

(b) CAV delay grouped by vehicle movements.
**Fig. 7.** Average travel time benefits of trajectory planning for a single CAV.

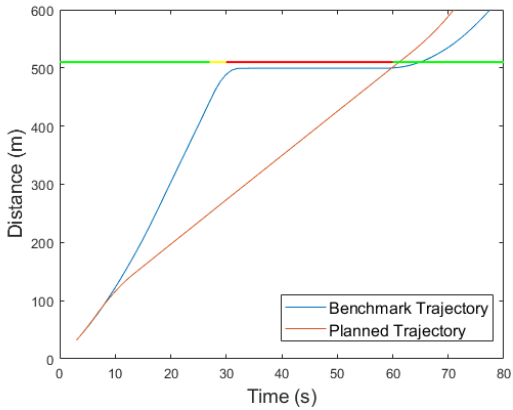 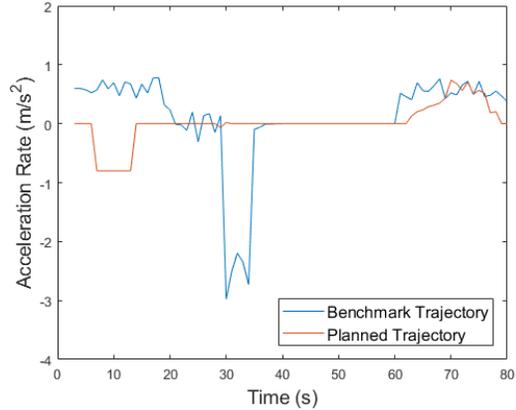

(a) Spatial-temporal trajectories.　　　　　　　　　　(b) Acceleration profiles.
**Fig. 8** Comparison of the planned trajectory and the benchmark trajectory.

**Fig. 9**(a) shows the average fuel economy of CAVs at various traffic demand levels. With increasing traffic, the fuel economy of CAVs decreases due to the congestion. At low demand levels (v/c <= 0.75), the optimized trajectories improve the fuel economy of CAVs by ~15% compared with the benchmark trajectories. At high demand levels (v/c >= 1), the improvement increases to ~30%. **Fig. 9**(b) shows the average fuel economy grouped by CAV movements. The benefits of the optimized trajectories are more remarkable for left-turning and through CAVs, which is ~20% at the low demand level (v/c = 0.25) and grows to ~ 35% at the high demand level (v/c=1.25). The main reason is that the optimized trajectories can smooth CAV trajectories, which helps reduce fuel consumption, as shown in **Fig. 8**(b).

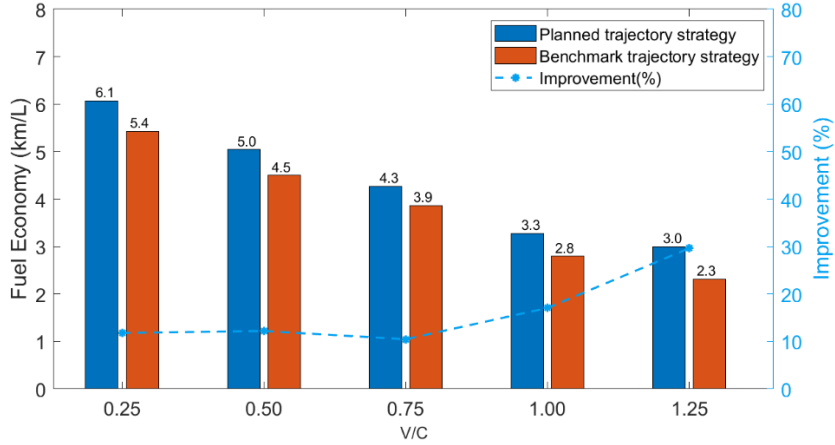
(a) CAV fuel economy.

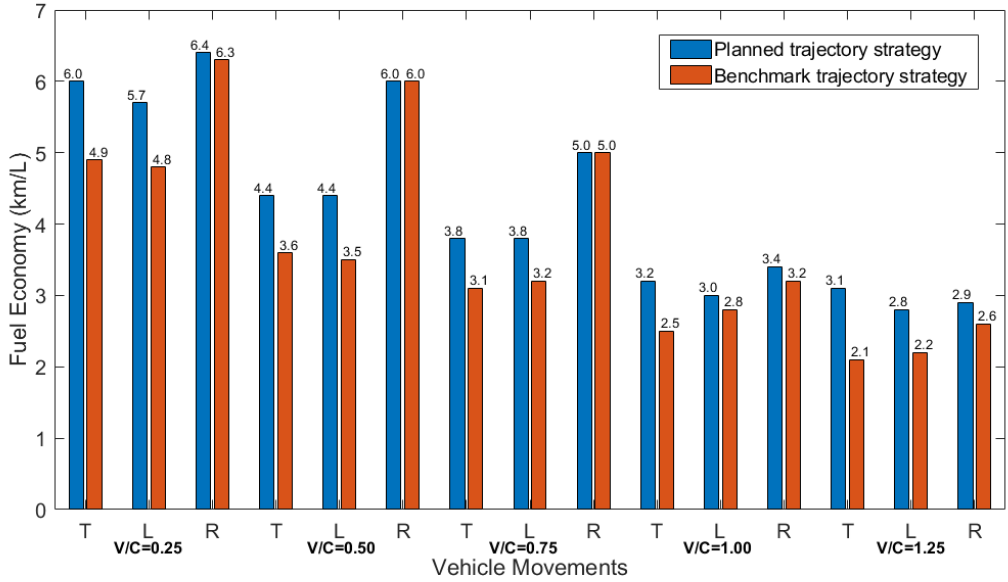
(b) CAV fuel economy grouped by vehicle movements.
**Fig. 9.** Average fuel economy benefits of trajectory planning for a single CAV.

**Fig. 10**(a) shows the average lane-changing numbers of CAVs at various traffic demand levels. At low demand levels (e.g., v/c=0.25), the optimized and the benchmark trajectories have comparable performance. Because lane-changing maneuvers of CAVs are scarcely affected by other vehicles in that case. In contrast, the advantages of the proposed trajectories over the benchmark trajectories are significant at high demand levels (e.g., v/c=1.25). The lane-changing number can be reduced by ~15% because of the reduced unnecessary lane-changing maneuvers. **Fig. 10**(b) shows the average lane-changing numbers grouped by CAV movements. It is observed that the lane-changing numbers of through CAVs are less than those of left- and right-turning CAVs. The reason is that through CAVs change lanes to the middle two through lanes more easily than left-/right-turning CAVs changing lanes to the left-/right-turning side lane. **Fig. 10**(b) also indicates that the optimized trajectories outperform the benchmark trajectories in most cases. Since the minimization of lane-changing numbers is the tertiary objective, CAVs may increase lane-changing numbers to reduce delay and improve the smoothness of longitudinal trajectories.

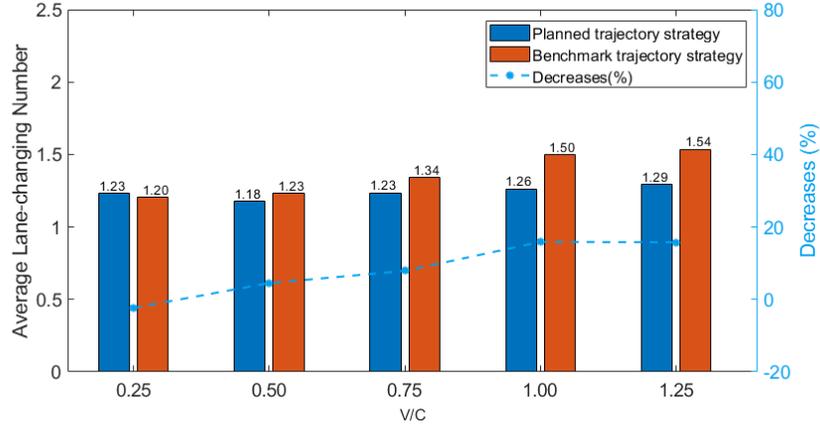

(a) CAV lane-changing numbers.

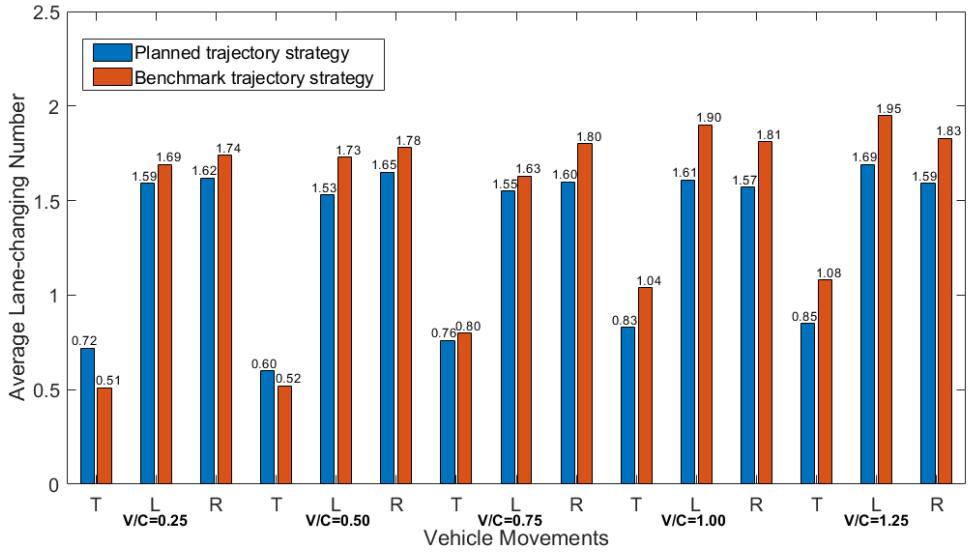

(b) CAV lane-changing numbers grouped by vehicle movements.
**Fig. 10.** Comparisons of the average lane-changing number for a single CAV.

### *5.2.3  Impacts on mixed traffic flow*

As described in Section 3, the trajectory planning for a CAV guarantees that the planned/predicted trajectories of its preceding vehicles are not affected for the concerns of fairness. In this way, the negative impacts of CAV trajectory planning on CHVs are expected to be alleviated. This section investigates the impacts of CAV trajectory planning on the mixed traffic flow and CHVs with different penetration rates of CAVs.

**Fig. 11**(a), (b), and (c) show the average delays of CAVs, CHVs, and the mixed traffic at the demand levels with v/c = 0.5, 1.0, and 1.25. With increasing demand, the average delays of CAVs, CHVs, and the mixed traffic rise significantly. With increasing penetration rates of CAVs, the three average delays decrease at each demand level. This indicates that the trajectory planning for CAVs has the potential to reduce the delays of CHVs and the mixed traffic when there are more CAVs in the mixed traffic. The delay reduction of CHVs is the most remarkable, which, for example, is ~4s when the CAV penetration rate increases over 70%. **Fig. 11**(d), (e), and (f) show the average delay reduction of CAVs, CHVs, and mixed traffic at different demand

levels compared with the benchmark cases in which no CAV trajectory planning is conducted. Although CAV trajectory planning is conducted in a decentralized way, it helps improve the operational performance of the mixed traffic with the CAV penetration rate higher than 20%. When the CAV penetration rate is high (e.g., 80%), the delay reduction of CHVs becomes significant, especially with over-saturated traffic (i.e., v/c=1.25). The reduced delay is more than 4 s. The reason is that CHVs could follow CAVs to form platoons to cross the intersection at high speeds without stops at the stop bar. **Fig. 12** shows the trajectories of CAVs and CHVs with the CAV penetration rate of 40% as an example. The trajectories segments in lane 2, which is a through lane, are marked in dark colors and those in other lanes are marked in light colors. The CHVs in the first cycle stops at the red light and start after the light turns green. As a result, these CHVs have the start-up lost time. In contrast, the CHVs in the third cycle follow leading CAVs to pass the stop bar at high speeds. Stops at the stop bar are avoided and the few seconds of the green time is utilized. Therefore, the trajectory planning of CAVs could reduce the delay of CHVs and the mixed traffic.

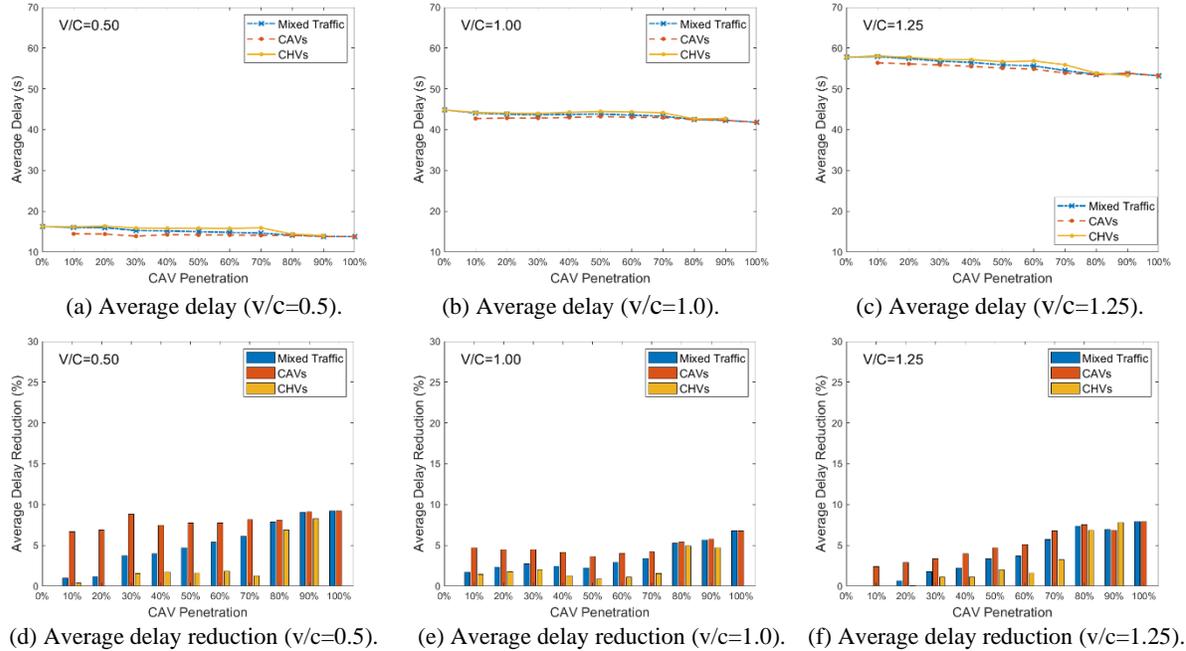

(a) Average delay (v/c=0.5).  (b) Average delay (v/c=1.0).  (c) Average delay (v/c=1.25).

(d) Average delay reduction (v/c=0.5).  (e) Average delay reduction (v/c=1.0).  (f) Average delay reduction (v/c=1.25).

**Fig. 11** Impacts on average delay of the CAV penetration at different demand levels.

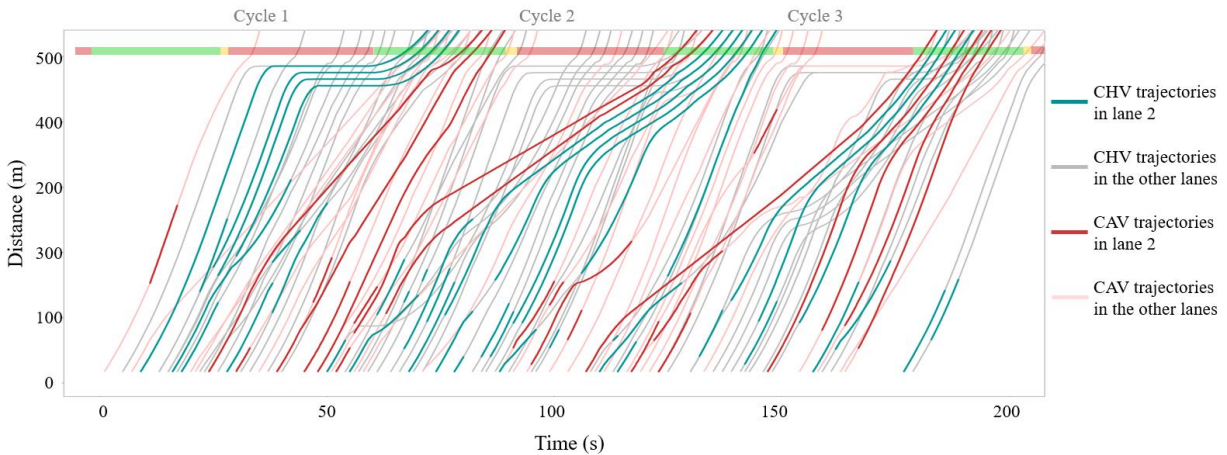

**Fig. 12** Spatial-temporal trajectories of CAVs and CHVs in the mixed traffic.

In addition, the capacity of the intersection could be enhanced due to the same reasons. **Fig. 13** shows the throughput of the approaching lanes with increasing demand. The CAV penetration rates of 0%, 50%, and 100% are tested. The capacity is reached when the throughput lines become flat. **Fig. 13** indicates that the capacity improvement is slight (~2%) when the CAV penetration rate increases from 0% to 50%. The maximum improvement is observed, which is ~6%, with the CAV penetration rate of 100%. These observations show that the trajectory planning of CAVs could improve the intersection capacity but a high penetration rate of CAVs is needed for a noticeable improvement when the signal timings are fixed.

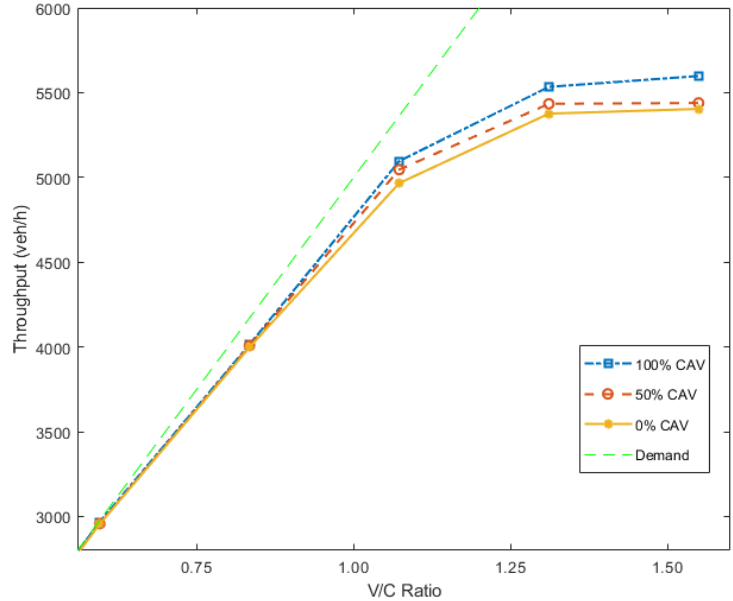

**Fig. 13** Throughput of approach lanes under different CAV penetration rates.

**Fig. 14**(a), (b), and (c) show the average fuel economy of CAVs, CHVs, and the mixed traffic at the demand levels with v/c = 0.5, 1.0, and 1.25. With increasing demand, the average fuel economy of CAVs, CHVs, and the mixed traffic decrease significantly. With increasing penetration rates of CAVs, the fuel economy increases, especially with high traffic. Because the trajectory planning of CAVs helps the mixed traffic to avoid stop-and-go motions and smooth longitudinal trajectories. Such benefits rise with increasing CAVs. **Fig. 15**(a) and (b) show the spatial distribution of the space mean speeds with v/c=0.5 and 1.25, respectively. When the CAV penetration rate is 0% (i.e., the fully CHV environment) with under-saturated traffic, the space mean speed increase gradually in the first 300 m of the control zone and then begins to drops to the lowest speed at the stop bar (500 m) as shown in **Fig. 15**(a). Because partial vehicles stop and queue at the locations between 300 m and 500 m during the red light. In contrast, the space mean speed begins to drop at 200 m with over-saturated traffic because of the congestion as shown in **Fig. 15**(b). The speed at the stop bar with over-saturated traffic is noticeably lower than that with under-saturated traffic. When the CAV penetration rate increases, the speed fluctuation over the control zone is reduced with both under- and over-saturated traffic, especially when the CAV penetration rate is over 60%. When the CAV penetration rate is 100%, the space mean speed over the whole control zone is relatively stable since the longitudinal trajectories of vehicles are smoothed by the trajectory planning. Vehicles' average speeds passing the stop bar are improved with increasing CAV penetration rates as well because of the avoidance of stop-and-go driving.

**Fig. 14** (d), (e), and (f) show the average fuel economy of CAVs, CHVs, and the mixed traffic at different demand levels compared with the benchmark cases in which no CAV trajectory planning is conducted. It can be observed that the trajectory planning of CAVs could also improve the fuel economy of CHVs and the mixed traffic although the fuel economy of CAVs is improved most remarkably. Generally, a higher penetration rate of CAVs leads to greater improvement of fuel economy. At high demand levels, CHVs are more likely to be affected by CAVs. As a result, a low penetration rate of CAVs is sufficient to improve the fuel economy of CHVs to a great extent at high demand levels while a high penetration rate of CAVs is needed at low demand levels. For example, the improvement reaches ~20% when the penetration rate is over 30% with the traffic of v/c-=1.25. But when the v/c=0.5, the improvement is less than 10% until the penetration rate increases to 70%.

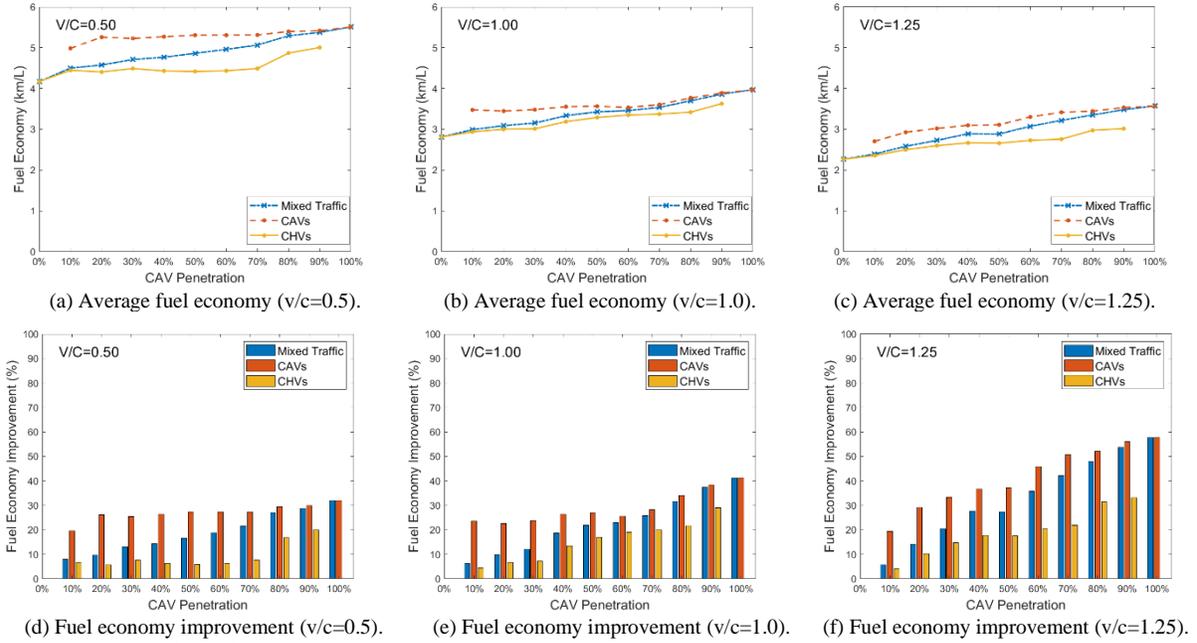

**Fig. 14** Impacts on fuel economy of the CAV penetration at different demand levels.

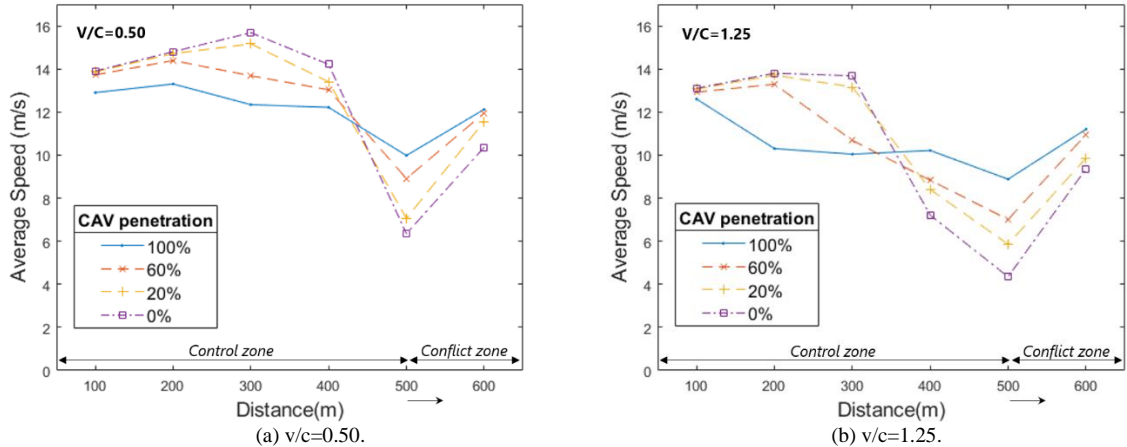

**Fig. 15** Spatial distribution of the space mean speed under different CAV penetration rates.

**Fig. 16**(a), (b), and (c) show the average lane-changing numbers of CAVs, CHVs, and the mixed traffic at the demand levels with v/c = 0.5, 1.0, and 1.25. Although the lane-changing numbers differ with different demand and different penetration rates of CAVs, the difference is insignificant ($\leq 0.2$). **Fig. 16**(d), (e), and (f) show the average lane-changing numbers of CAVs, CHVs, and the mixed traffic at different demand levels compared with the benchmark cases in which no CAV trajectory planning is conducted. The trajectory planning for CAVs could slightly increase the lane-changing frequency of CHVs when the CAV penetration rate is less than 50%, which is consist with previous studies (Zhong et.al., 2020). CAVs may slow down in the upstream approach lanes to avoid stops at stop bars. Consequently, the following CHVs may choose to change lanes. In most cases, the average lane changing number of CAVs is less than that of CHVs with CAV trajectory planning. Further, the reduction of the lane changing number of CAVs is remarkable, which is ~10%, at the high demand level (v/c = 1.25) compared with the benchmark.

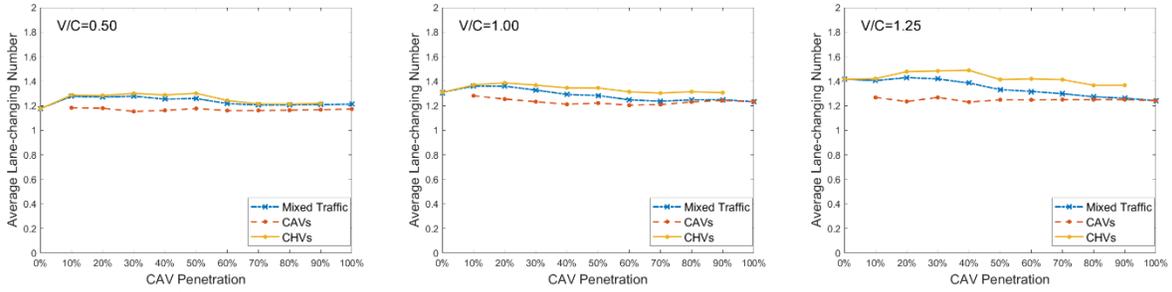

(a) Lane-changing number (v/c=0.5). (b) Lane-changing number (v/c=1.0). (c) Lane-changing number (v/c=1.25).

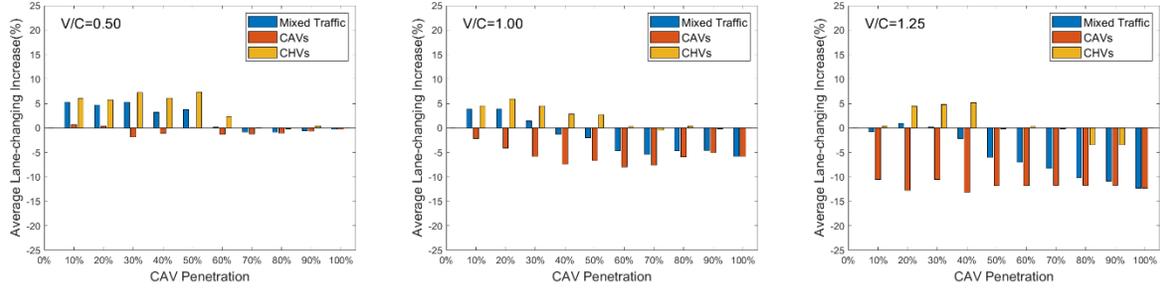

(d) Lane-changing number increase (v/c=0.5). (e) Lane-changing number increase (v/c=1.0). (f) Lane-changing number increase (v/c=1.25).

**Fig. 16** Impacts on lane-changing number of CAV penetration under different demand levels.

*5.3 Sensitivity analysis*

The length of the control zone plays an important role in the trajectory planning of CAVs in the mixed traffic environment. A larger length enlarges the solution space, which may lead to better operational performances of the intersection traffic, but at the cost of the computational burden. **Fig. 17** shows the impacts of the control zone length on the performance of the CAV trajectory planning model when the demand level is v/c=0.75 and the CAV penetration rate is 40%. The dashed lines illustrate the benchmark cases without CAV trajectory planning. As shown in **Fig. 17**(a), there is a noticeable reduction of delay when the length of the control zone increases to 100 m. When the length increases further, the delay reduction is insignificant. That is, the control zone of 100 m is sufficient for CAV trajectory planning to avoid stops at the stop bar and cross the intersection at high speeds. In contrast, the increase of the fuel economy slows down remarkably when the control zone length is larger than 200 m as shown in **Fig. 17**(b). This indicates the smoothness of longitudinal trajectories required a larger control zone. The impacts of the control zone length on the lane-hanging number are unstable as shown in **Fig. 17**(c). In conclusion, it is suggested that the control zone length should be at least longer than 200 m.

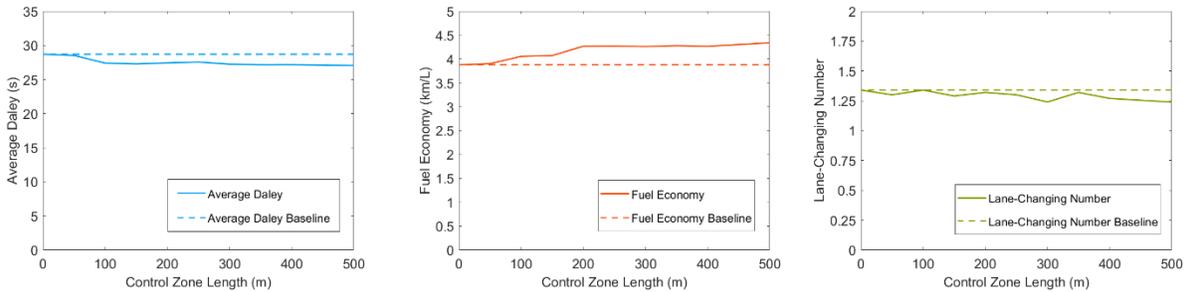

(a) Average delay. (b) Fuel economy. (c) Lane-changing number.
**Fig. 17** Sensitivity analysis on the impact of the control zone length.

# 6  Conclusions and recommendation

This study proposes a trajectory planning model for CAVs in a decentralized way under the mixed traffic environment which consists of CHVs and CAVs. CAV trajectories are planned one by one according to their longitudinal locations in the approach lanes. The trajectory planning of a single CAV is decomposed into a lateral lane-changing strategy and a longitudinal acceleration profile. A bi-level optimization model is then built based on signal timing plans and planned/predicted trajectories of CAVs and CHVs. The upper-level model optimizes the lane-changing strategies in which the concepts of LCGs are proposed. The lower-level model optimizes the acceleration profiles which guarantees the planned/predicted trajectories of preceding CAVs and CHVs are not affected for the concerns of fairness. An LCST and a PMCTS algorithm are designed to solve the bi-level model. Parallel computing can be used to improve the computational efficiency. A rolling horizon scheme is applied for the dynamic implementation of the proposed model with time-varying traffic condition. The numerical studies validate the advantages of the proposed trajectory planning model compared with the benchmark cases without CAV trajectory planning. The simulation results show that CAV fuel consumption and lane-changing numbers can be reduced noticeably, especially with high traffic demand. CAV delay is reduced by ~2 s on average, which is limited due to the fixed signal timing plans. Although CAV trajectories are planned in a decentralized way, the delay and the fuel consumption of CHVs and the mixed traffic is reduced as well, especially with high penetration rates of CAVs. However, the lane-changing numbers of CHVs and the mixed traffic may be increased with low penetration rates of CAVs. The sensitivity analysis shows that the control zone length of 200 m is sufficient from the perspective of the performance of the proposed CAV trajectory planning model.

In this study, signal timing plans at the intersection are fixed. This limits the improvement of the operational performance of the mixed traffic. The integration of signal timing optimization and CAV trajectory planning under the mixed traffic environment is planned in our following research. It is another challenge to take regular vehicles into consideration that are neither observable nor controllable. The difficulty lies in the estimation and prediction of the states of regular vehicles. Furthermore, we plan to extend the CAV trajectory planning model from the intersection level to the network level. The macroscopic routing and the microscopic trajectory planning can be combined.

## Acknowledgments

This paper is supported by National Key R&D Program of China (No. 2018YFB1600600), the National Natural Science Foundation of China (No. 61903276), and Shanghai Sailing Program (No. 19YF1451600).